# Kinetic vs. multi-fluid approach for interstellar neutrals in the heliosphere: exploration of the interstellar magnetic field effects

Fathallah Alouani-Bibi[1], Merav Opher[1,2], Dimitry Alexashov[3], Vladislav Izmodenov[3,4], Gabor Toth[5]

[1] Physics and Astronomy Department, George Mason University, USA.

[2] Astronomy Department, Boston University, Boston, USA

[3] Space Research Institute (IKI) and Institute for Problems in Mechanics,

Russian Academy of Science, Moscow, Russia

[4] Lomonosov Moscow State University, Moscow Russia

[5] Center for Space Environment Modeling, University of Michigan, USA

alouani@physics.gmu.edu

mopher@bu.edu

## Abstract

We present a new 3d self-consistent two-component (plasma and neutral hydrogen) model of the solar wind interaction with the local interstellar medium (*LISM*). This model (K-MHD) combines the MHD treatment of the solar wind and the ionized *LISM* component, with a kinetic model of neutral interstellar hydrogen (*LISH*). The local interstellar magnetic field ($B_{LISM}$) intensity and orientation are chosen based on an early analysis of the heliosheath flows (Opher *et al*. 2009). The properties of the plasma and neutrals obtained using the (K-MHD) model are compared to previous multi-fluid (Opher *et al*. 2009) and kinetic models (Izmodenov *et al*. 2005). The new treatment of *LISH* revealed important changes in the heliospheric properties not captures by the multi-fluid model. These include a decrease in the heliocentric distance to the termination shock (TS), a thinner heliosheath and a reduced deflection angle (θ) of the heliosheath flows. The asymmetry of the termination shock, however, seems to be unchanged by the kinetic aspect of the *LISH*.

**1. Introduction**

The interaction of the supersonic solar wind with the local interstellar medium (*LISM*) as the Sun travels through the galaxy results in the formation of the heliosphere in which our solar system is imbedded (Axford 1972, Baranov 1990, Zank 1999, Pauls *et al*. 1995, Izmodenov & Kallenbach, 2006). Within the termination shock (TS), the closest of the three heliospheric boundaries to the Sun, the solar wind is supersonic with latitudinal variations dependent on the solar cycle (McComas *et al*. 2006; Bzowski 2008, Izmodenov & Malama 2004). At the TS, the solar wind is decelerated to subsonic speeds and heated. The *LISM* and solar wind reach a pressure balance at the heliopause (HP), the second heliospheric boundary. Outside the heliopause, the *LISM* plasma and the frozen-in magnetic field drape over this boundary. The *LISM* neutrals on the other hand, not affected by the electromagnetic force, stream relatively freely through these heliospheric boundaries. Neutrals, mainly hydrogen, do however interact with the solar wind and *LISM* plasma through a resonant charge-exchange (Lindsay & Stebbings 2005). This leads to a deceleration and a deflection of the interstellar neutrals outside the HP resulting in an enhanced density structure, "the hydrogen wall" (Baranov & Malama 1993; Linsky & Wood 1996). The third heliospheric boundary, the bow shock (BS) is a result of a deceleration of the *LISM* outside the HP. For a strong interstellar magnetic field, however, as we now believe a bow shock is absent (Izmodenov *et al*., 2009, Opher *et al*. 2009).

Voyager 1 (V1) and 2 (V2) recently reached the TS (Stone et al 2005, 2008), giving a new and unique perspective of that region. One of the major findings of these two missions was the so-called

"north-south" asymmetry, referring to a 10 AU difference in the locations of the TS at Voyager 1 and 2. This asymmetry is believed to originate outside the HP and is most likely due to an asymmetric magnetic pressure (Opher *et al*. 2006). Some of this asymmetry, we believe to a lesser extent, is due to variations in the solar wind ram pressure during the period between V1 and V2 crossings.

The local interstellar magnetic field ($B_{LISM}$) is the least known parameter of the *LISM*. A strong constraint on the orientation of $B_{LISM}$ was established by the SOHO/SWAN observations (Lallement *et al*. 2005, 2010). These measurements showed a deflection of the hydrogen flow by ~4° relative to the helium flow. The later is assumed to be unaffected by the crossing of the heliosphere owing to its negligible charge-exchange coupling with the heliospheric plasma. Most of the deflection of hydrogen takes place outside the heliopause. The SOHO/SWAN measured deflection has been interpreted as due to the local interstellar magnetic field (Izmodenov *et al*. 2005, Pogorelov et al. 2008, Alouani-Bibi *et al*. 2010).

Extended numerical modeling has been dedicated to understanding the heliosphere formation and dynamic and reproducing some of the observations (Baranov & Malama 1993, Zank *et al*. 1999; Linde *et al*. 1998; Myasnikov *et al*. 2000; Opher *et al*. 2006, 2007, 2009, Izmodenov *et al*. 2005, 2009, Malama *et al*., 2006, Ratkiewicz & Grygorczuk 2008, Pogorelov *et al*. 2008, 2009). These studies underscored the key role that neutral *LISM*, mainly hydrogen, plays in the heliosphere. Neutral hydrogen dominates the ionized component of the *LISM* in terms of number density ($n_H/n_p$ ~ 3). Therefore, the heliospheric asymmetry (Izmodenov *et al*., 2005; Opher *et al*. 2006, 2007; Pogorelov *et al*. 2008, 2009), the heliosheath size and the inferred properties of $B_{LISM}$ (Pogorelov *et al*. 2008, 2009; Opher *et al*. 2009; Izmodenov 2009) are expected to be strongly affected by the neutral *LISM* component.

In the heliosphere, the Knudsen number (the ratio of the charge-exchange mean free path to the scale length of the system) for neutral hydrogen is greater than 1 (e.g. Izmodenov *et al*. 2000). A full kinetic treatment of neutrals is therefore needed. Solving the kinetic equation is computationally challenging and time consuming, and most often approximations are used. One of these approximations is the multi-Maxwellian model, whereby neutrals are represented by multi-fluid species. Each specie reflects the peculiar thermodynamic properties of the plasma between adjacent heliospheric boundaries (Zank *et al*. 1996; Alexashov & Izmodenov *et al*. 2005, Opher *et al*. 2009). Notwithstanding the fact that the multi-fluid model is a good approximation for describing the main features of the heliosphere, it has however some limitations (Alexashov & Izmodenov 2005) and most often the accuracy of the model depends on the number of neutral species considered. Kinetic-hydrodynamic coupling procedures, i.e. kinetic neutrals and hydrodynamic plasma, were introduced two decades ago, and have been successfully applied to different geometries and parameters of the solar wind and the *LISM* (Malama 1991, Baranov & Malama 1993, Müller *et al*. 2000, Izmodenov & Malama 2004; Izmodenov *et al*. 2005, Heerikhuisen *et al.* 2006).

Here we extend our 3d MHD multi-fluid model (MF-MHD) (Opher *et al*. 2009) to include a kinetic treatment of neutrals. The kinetic model has been developed by the Moscow group (Malama 1991; Baranov & Malama 1993; Izmodenov & Malama 2004; Izmodenov *et al*. 2005). In this new framework, we analyze the effect of neutral *LISM* on the location and the asymmetry of the heliospheric boundaries as well as on the properties of the heliosheath flows. We compare the results from the new model (K-MHD) and previous data using multi-fluid (MF-MHD) (Opher *et al*. 2009) and kinetic (A-I) (Izmodenov *et al.* 2005, Alexashov & Izmodenov 2005) models.

The paper is structured as follows: Section 2 describes the two models used for neutrals, multi-fluid and kinetic, and the MHD-kinetic coupling procedure. The boundary values for the solar wind and the

*LISM,* the geometry of the problem are laid out in section 3. The results and discussion are presented in section 4. Conclusions are in section 5.

## 2. Models

We use a 3d ideal MHD single ion fluid model to describe both the solar wind and the ionized component of the *LISM*. This is done using BATS-R-US code (Powell *et al*. 1999, Toth *et al.* 2005). The governing equations for the ionized component and magnetic field are:

$$\frac{\partial \rho}{\partial t} + \vec{\nabla} \cdot \left( \rho \vec{u} \right) = 0 \tag{1}$$

$$\frac{\partial (\rho \vec{u})}{\partial t} + \vec{\nabla} \cdot (\rho \vec{u} \otimes \vec{u}) + \vec{\nabla} \cdot \left( p^* \overset{\leftrightarrow}{I} - \frac{(\vec{B} \otimes \vec{B})}{\mu_0} \right) = \vec{S}_M \tag{2}$$

$$\frac{\partial E}{\partial t} + \vec{\nabla} \cdot (E \vec{u}) + \vec{\nabla} \cdot \left( p^* \vec{u} - \frac{(\vec{u} \cdot \vec{B}) \vec{B}}{\mu_0} \right) = S_E \tag{3}$$

$$\frac{\partial \vec{B}}{\partial t} = \vec{\nabla} \times (\vec{u} \times \vec{B}) \tag{4}$$

$$E = \frac{1}{2} \rho u^2 + \frac{p}{(\gamma - 1)} + \frac{B^2}{2\mu_0} \tag{5}$$

$$p^* = p + \frac{B^2}{2\mu_0} \tag{6}$$

The terms $S_M$ and $S_E$ are the momentum and energy source terms due to the charge-exchange between neutrals and ions. These terms are derived in (McNutt *et al*.1998) for a single neutral population. They represent the combined contribution to the ion's momentum and energy by each of the neutral species. Here we consider 4 neutral populations.

$$\vec{S}_M = \sum_{j=1}^{4} \rho_p \nu_{H(j)} \left( \vec{U}_{H(j)} - \vec{U}_p \right) \tag{7}$$

$$S_E = \sum_{j=1}^{4} \rho_p \, \nu_{H(j)} \left[ \left( \frac{U_{H(j)}^2 - U_p^2}{2} \right) + \frac{U_{H(j)}^*}{U_{M,H(j)}^*} \left( V_{Th,H(j)}^2 - V_{Th,p}^2 \right) \right] \tag{8}$$

The index "j" refers to the neutral specie, $\nu_{H(j)} = n_H \, U_{M,H(j)}^* \, \sigma_{ChEx}$ and $\sigma_{ChEx}$ are the charge-exchange frequency and cross section (Lindsay & Stebbings 2005) respectively. The bulk and the thermal speeds of neutral specie "j" and ions are $(U_{H(j)}, U_p)$ and $(V_{Th,H(j)}, V_{Th,p})$ respectively. The terms $(U_{H(j)}^*, U_{M,H(j)}^*)$ are the characteristic speeds at which the charge-exchange cross section is evaluated (see Eqs. (62)-(64) in McNutt *et al.* 1998). The expression $\sigma_{ChEx}(U_{M,H(j)}^*)$ is used to calculate the momentum and energy, while $\sigma_{ChEx}(U_{H(j)}^*)$ is used for the mass conservation equation for neutrals.

$$U_{H(j)}^* = \sqrt{\left| U_{H(j)} - U_p \right|^2 + \frac{4}{\pi} \left( V_{th,H(j)}^2 + V_{th,p}^2 \right)} \tag{9}$$

$$U_{M,H(j)}^* = \sqrt{\left| U_{H(j)} - U_p \right|^2 + \frac{64}{9\pi} \left( V_{th,H(j)}^2 + V_{th,p}^2 \right)} \tag{10}$$

There is no density source term in Eq. 1, as the ions number density in a single ion fluid model is unchanged under the charge-exchange process. In equations (2-3) the radiation pressure and the gravity are assumed to perfectly cancel each other out. Ionization processes such as photo-ionization and electron-impact ionization are also neglected. These processes play a much lesser role than the charge-exchange at larger radii ($R>30AU$). Electron impact ionization may play some role in the heliosheath. Unfortunately, a detailed description of the electron velocity distribution function is needed to correctly calculate the rates of this process (see Allais *et al.* 2005), which is beyond the scope of this paper.

For the neutral component, hydrogen, we use both the multi-fluid and the kinetic models. The multi-fluid treatment of neutrals assumes 4 neutral populations each reflecting the properties of the plasma between the different heliospheric boundaries (see Figures 1a, 2a-b). An extended study of the heliosheath flows with the multi-fluid model and comparisons with Voyager data can be found in (Opher *et al.* 2009).

The governing equation for neutrals in the multi-fluid model, are:

$$\frac{\partial \rho_{H(j)}}{\partial t} + \vec{\nabla} \cdot \left( \rho_{H(j)} \vec{w} \right) = S_{\rho,H(j)} \tag{11}$$

$$\frac{\partial (\rho_{H(j)} \vec{w})}{\partial t} + \vec{\nabla} \cdot (\rho_{H(j)} \vec{w} \otimes \vec{w}) + \vec{\nabla} \cdot \left( p_{H(j)} \overleftrightarrow{I} \right) = \vec{S}_{M,H(j)} \tag{12}$$

$$\frac{\partial E_{H(j)}}{\partial t} + \vec{\nabla} \cdot (E_{H(j)} \vec{w}) + \vec{\nabla} \cdot \left( p_{H(j)} \vec{w} \right) = S_{E,H(j)} \tag{13}$$

$$\chi = \begin{cases} 1 & \text{if } \vec{r}_{H(j)} \in Zone_{H(j)} \\ 0 & \text{otherwise} \end{cases} \tag{14}$$

$$S_{\rho,H(i)} = \chi \left( \sum_{j=1}^{4} \rho_p \nu_{H(j)} \right) - \rho_p \nu_{H(i)} \tag{15}$$

$$\vec{S}_{M,H(i)} = \chi \left( \sum_{j=1}^{4} \rho_p \nu_{H(j)} \vec{U}_p \right) - \rho_p \nu_{H(i)} \vec{U}_{H(i)} \tag{16}$$

$$S_{E,H(i)} = \chi \left( \sum_{j=1}^{4} \rho_p \nu_{H(j)} \left[ \frac{U_p^2}{2} + \frac{U_{H(j)}^*}{U_{M,H(j)}^*} V_{Th,p}^2 \right] \right) - \rho_p \nu_{H(i)} \left[ \frac{U_{H(i)}^2}{2} + \frac{U_{H(i)}^*}{U_{M,H(i)}^*} V_{Th,H(i)}^2 \right] \tag{17}$$

where "i" is the index of the neutral specie. The function $\chi$ takes the value 1 when the neutral population (i) has its position vector inside its predefined source region (see below for the definition of these regions, Figures 1a, 2a). The terms ($S_{M,H(i)}$, $S_{E,H(i)}$) are the momentum and energy source terms for the neutral population (i) due to the charge-exchange with ions. Under this formulation, where each population has a predefined source region, the number density of each population is not a conserved quantity while the total number density of all neutral species is.

To define the source region of each neutral population, a priori assumption regarding the location of the TS and HP is needed. This assumption is based on the behavior of the plasma (density, temperature, bulk speed and magnetic field) throughout the heliosphere. In other words, the source regions for neutrals are chosen based on MHD criteria.

The source region for the primary *LISM* neutrals (population 4), farther from the heliopause at the undisturbed *LISM* or beyond the bow shock in the case of a low intensity $B_{LISM}$, is defined as having magneto-sonic Mach number $M_{mag}$ and flow speed $U$ as ($M_{mag} > 0.5$) and ($U < 100$ km/s) respectively. The region between the bow shock and the HP, where the secondary *LISM* neutrals (population 1) are produced through charge-exchange with the *LISM* plasma, satisfy ($M_{mag} < 0.5$, $U < 100$ km/s and $T < 10^5$ K). Population 2 is produced in the heliosheath, by charge-exchange between the *LISM* neutrals and the shocked solar wind. This region is defined with characteristic sub-sonic speed with $M_{mag} < 0.5$ and temperature $T > 10^5$ K. Population 3 is generated within the TS as a result of charge-exchange between the *LISM* neutrals the supersonic solar wind.

The steady state density distribution of these neutral populations is shown in Figures 2a-b and Figures 3a-b. Figure 2a, portrays a 2d map of the density of each population, while Figure 2b show the sum of all populations. The "hydrogen wall" structure is clearly visible in Figure 2b. Radial density profiles, taken along the X axis, in the upwind *LISM* flow direction, are shown in Figures 3a-b.

The kinetic model of neutral hydrogen is based on the solution of the kinetic equation using the Monte-Carlo method. Details of the model are presented in (Malama 1991; Baranov & Malama 1993; Izmodenov & Malama 2004; Izmodenov *et al*. 2005). The Boltzmann charge-exchange collision operator for neutrals, including the production and the ionization terms, is:

$$S = -f_H(\vec{w}_H)\int|\vec{w}_H - \vec{w}_P|\sigma_{ChEx}f_p(\vec{w}_P)d\vec{w}_P + f_p(\vec{w}_H)\int|\vec{u}_H - \vec{w}_H|\sigma_{ChEx}f_p(\vec{u}_H)d\vec{u}_H - (v_{ph} + v_{imp})f_H(\vec{w}_H) \quad (18)$$

where ($f_H$, $f_p$) are the velocity distribution functions of neutrals and protons respectively. In this model the protons are represented by a Maxwellian distribution function, defined by the local plasma temperature, density and bulk speed ($n_p$, $T_p$, $U_p$). These plasma parameters are taken as input from the MHD calculations at each kinetic-MHD iteration. The multi-component nature of the ionized component is not considered in this paper (see, Malama *et al*. 2006; Chalov *et al*. 2010). Photoionization and electron impact ionization terms in (Eq. 18), i.e. $(\nu_{ph}, \nu_{imp})$, are not taken into account in the kinetic model, as is the case in the mass conservation equation (Eq.1) for the MHD model.

The steady state solution with the K-MHD model is reached after a series of iterative steps. The initial phase, step 1, of this iterative process is achieving a steady state solution for a given set of initial parameters of the *LISM* and the solar wind. The steady state is attained using a 3d MHD model with a multi-fluid (4 neutrals populations) description of neutrals. Both the *LISM* and the interplanetary magnetic fields are considered in our calculations. A Monte-Carlo simulation of neutrals is carried out in step 2, using an updated proton distribution function in (Eq.18). The distribution function is updated using local plasma properties ($n_p$, $T_p$, $U_p$) taken from the MHD simulation. Thereafter, the generated kinetic neutrals quantities ($n_H$, $T_H$, $U_H$), are used in step 3 to calculate the momentum and energy source terms ($S_M$, $S_E$) for the ions. Therefore (Eqs. 7-8) is used from this point on in the iterative process, with a single kinetic population of neutrals. Therefore (Eqs. 11-18) are no longer updated. Step 3, is run for 500 time steps, before the output ($n_p$, $T_p$, $U_p$) is sent back for step 2. Steps 2 and 3 are repeated until a steady state is achieved between the kinetic neutrals and the MHD plasma. The coupling procedure converges quickly (after $4^{th}$ iteration). We use the same stationary adaptive mesh refinement (AMR) grid for the MHD and kinetic calculations. This is done to avoid any incremental interpolation errors due to the iterative process. The grid has been optimized to capture the main heliospheric features and the heliospheric boundaries.

## 3. Solar wind and *LISM* boundary values

We consider a fully ionized solar wind plasma with number density, temperature and flow speed given by $n_{swp} = 8.7 \times 10^{-3} cm^{-3}$, $T_{swp} = 10^5 K$, $V_{swp} = 417 \, km/s$ (OMNI solar data, http://omniweb.gsfc.nasa.gov/). These values are assumed at the inner-boundary, located at 30 AU from the sun. The interplanetary magnetic field is given by the Parker solution (Parker 1961) at the inner boundary.

The solar wind parameters at 30 AU, inner-boundary for the MHD model, are matched to those used in the kinetic model at 30 AU. The kinetic, Monte-Carlo, simulations have an inner-boundary at 1 AU. The solar wind parameters are therefore extrapolated from 30 AU to the 1 AU. The extrapolation assumes an adiabatic solar wind with a constant speed: $n_{swp} \sim n_0 / r^2$; $p_{swp} \sim p_0 / r^{2\gamma}$, $V_{swp} = 417 \, km/s$.

At the outer boundary the local interstellar medium is considered with: $n_{LISM,p} = 0.06 \, cm^{-3}$, $n_{LISM,H} = 0.18 \, cm^{-3}$, $T_{LISM} = 6530 K$, $V_{LISM} = 26.4 \, km/s$ (Witte 2004; Möbius *et al.* 2004, Gloeckler *et al.* 2004). The intensity of the magnetic field carried by the *LISM* plasma is chosen as 4.4 µG, while the orientation of the field is varied. We define the orientation of the local interstellar magnetic field vector by two angles (α, β), where α is the angle between $B_{LISM}$ and interstellar flow velocity $V_{LISM}$, and β is the angle between the *BV* plane (the plane containing both $B_{LISM}$ and $V_{LISM}$) and the solar equatorial plane (see Figure 1b). The values assigned to these angles are (α=20°, 30°) and (β=60°, 80°, 90°) respectively. These are chosen based on a previous (MF-MHD) analysis of the heliosheath flows (Opher *et al.* 2009). The heliospheric asymmetry and the position of the TS are well reproduced with these orientations for a field intensity of 4.4 µG (Opher *et al.* 2009). The case of (β = 60°) was however less successful in describing the heliosheath flows. Nevertheless, it is important to mention that the value (β) that most closely matches the inferred orientation of $B_{LISM}$ from the IBEX data

(McComas *et al*. 2009) is (β = 60°). This value has also been reported by (Pogorelov *et al*. 2008, Alouani-Bibi *et al*. 2010) to explain the hydrogen deflection in respect to helium (Lallement *et al*. 2005, 2010).

We adopt here the same convention for (α, β) as in (Opher *et al*. 2009). The angle α is equal to zero for parallel $B_{LISM}$ and $V_{LISM}$ vectors, and increases from $V_{LISM}$ counterclockwise toward +Z. The angle β is equal to zero at +Y and increases counterclockwise from +Y to +Z. The coordinate system is such that the Z axis coincides with the solar rotation axis, and the X axis lays on the plane spanned by the interstellar flow velocity vector $V_{LISM}$ and the Z axis. The $V_{LISM}$ is at ~-5° in respect to the X axis. The Y axis completes the orthogonal coordinate system. In this coordinate system, the interstellar flow velocity vector has coordinates of (26.3, 0, -2.3) km/s.

We use an adaptive mesh refinement (AMR) grid with 11 levels of refinement (see Toth *et al*. 2005). The grid spans from -1500 AU to 1500 AU in each of the (X, Y, Z) direction. The spatial resolution varies from ~0.7 AU near the inner-boundary to ~100 AU at the outer-boundaries.

## 4. Results and discussion

We consider 6 different configurations of $B_{LISM}$, each defined by a unique combination of (α, β). The values chosen for (α, β) are (α=20°, 30°) and (β=60°, 80°, 90°) respectively. The $B_{LISM}$ intensity is kept constant at 4.4 µG. The plasma properties described below are the number density, temperature, flow speed and magnetic field.

The magnetic field intensity in the XZ plane is shown in Figure 4a-b, for both the MF-MHD and the K-MHD models. For this particular case, the intensity and the direction of $B_{LISM}$ are set to 4.4 µG and (α=20°, β=80°) respectively. For this configuration, the same is in fact true for all the other configurations considered in this paper, the current sheet is directed northward. The 2d map (Figures

4a-b), shows variation in the field intensity between the two models used for neutrals. This intensity increase (Figure 4b) outside the HP is due to an unequal inward shift of the locations of the HP and the TS when the kinetic model is used to describe neutrals, as we will show later in the text. As a result, a net compression outside the HP takes place thus leading to an increase in the field intensity.

The protons number density profiles along Voyager 1 (V1) and Voyager 2 (V2) trajectories are presented in (Figure 5). These profiles show a systematic decrease in the heliocentric distance to the TS and HP for the K-MHD model as compared to the MF-MHD model. This reduction is mostly sensitive to (α), the angle between $B_{LISM}$ and $V_{LISM}$.

Magnetized shocks in a heliospheric environment have in general a characteristic thickness comparable to the thermal proton's gyroradius. This quantity is many orders of magnitude smaller than our grid resolution. We therefore introduce a concept of "mean value" to account for the illusiveness of the precise location of the heliospheric discontinuities in our calculations. The mean values for both TS and HP locations are calculated using the density and the velocity profiles. This is done by simply averaging the immediate upstream and downstream locations (see Table 1).

The mean heliocentric distance $<R_{TS}>$ to the TS is estimated for different values of (α , β) (see Table 1). For α=20°, varying β by 30° (Δβ=30°) induces a change in $<R_{TS}>$ of ~4 AU for the kinetic model and ~5.7 AU for the multi-fluid model. Similar estimates made for a constant β and a varying α, showed a larger variation in $<R_{TS}>$. For β=60°, varying α by 10° (Δα=10°) generated changes in $<R_{TS}>$ by ~16.3 AU and ~17.3 AU for the kinetic and multi-fluid models respectively. The relative decrease in the mean heliocentric distance to the termination shock, $<R_{TS}>$, when neutrals are kinetic has also been reported in other works (Pogorelov *et al*. 2008, Alexashov & Izmodenov 2005). Similar behavior is captured by the (A-I) model (see Figure 9) (Izmodenov *et al*. 2005, Alexashov & Izmodenov 2005). This reduction in $<R_{TS}>$ is a consequence of the solar win ram pressure decrease, as seen in velocity profiles (Figures 7a-b). This is due to an increase of the filtration factor (ratio of hydrogen number densities inside to the outside of the HP) in the kinetic model.

The density profiles (Figure 5) also show a net decrease in the mean heliosheath thickness $<L_{HS}>$ for the kinetic model, for all the considered cases of (α, β). The only exception has been the case of (α=30°, β=80°), for which both models, i.e. kinetic and multi-fluid, gave similar values (~ 60 AU). The heliosheath thickness has the same trend as the TS, in that it has a greater dependence on (α) than on (β). Moreover, both $<L_{HS}>$ and $<R_{TS}>$ are inversely proportional to (α). This is simply the result that for smaller (α) the (V x B) force term outside the HP is small, thus a more expanded heliosheath. It is also worth mentioning that dependencies on (β) are neither uniform nor preserved for different (α).

The temperature profiles (Figure 6) show similar features at the heliospheric boundaries as for the density profiles. The heliosheath plasma is slightly cooler when neutrals are kinetic, which is the result of the increased rate of charge-exchange with the shock heated solar wind. This is also confirmed by the strong deceleration of the heliosheath plasma in the kinetic model (see Figures 7a-b). The difference in the velocity profiles between the two models, for a given $B_{LISM}$ configuration, extends to the region within the TS, especially for the case (α=20°, β=60°). The temperature increase and velocity decrease within the TS boundary both point to a possible secondary charge-exchange process. Whereby some of the neutrals that charge-exchanged in the heliosheath went through another charge-exchange process with the supersonic solar wind component. Causing therefore not only the slowing down but a net heating of the solar wind.

The asymmetry of the TS between the V1 and the V2 crossing sites ($\Delta R_{TS\_V1V2}$), is only slightly affected by the treatment of neutrals ( $\Delta(\Delta R_{TS\_V1V2})_{Kinetic\_Multi-Fluid} \leq 1AU$ ). The exception being (α=20°, β=60°), in this case ( $\Delta(\Delta R_{TS\_V1V2})_{Kinetic\_Multi-Fluid} \sim 1.9\ AU$ ) (see Table 1). On average, accounting for all possible combinations of (α, β) in the chosen domain (α = 20°, 30° and β = 60°, 80°, 90°), the TS asymmetry between V1 and V2 crossing sites ( $<\Delta R_{TS\_V1V2}>$ ) is ~6.9 AU for the kinetic model and ~7AU for the multi-fluid.

The deflection angle θ of the heliosheath flows, ($\theta = \tan^{-1}(V_N/V_T)$), is shown in (Table 1). This angle expresses the ratio of the normal ($V_N$) and the tangential ($V_T$) velocity components downstream of the TS along V2 trajectory. These velocity components ($V_N$, $V_T$) (Figure 7a-b) are calculated in the *RTN* spacecraft frame. Neutrals have a strong effect on the deflection angle, with the general trend of decreasing |θ| for the K-MHD model as compared to the MF-MHD model. The largest differences (Δθ = $\theta_{Kinetic} - \theta_{MF}$) between the two models are attained for cases where (α = 20°). These differences vary from ~4.5° to ~10°, with the maximum at (α = 20°, β = 60°). For the cases where (α = 30°), Δθ varied between ~-1.2° to ~4.5°, where (α=30°, β=80°) is the only case where the kinetic model showed a higher value of |θ|. The detailed dependence of the deflection angle on the local interstellar magnetic field intensity and direction is shown in (Opher *et al*. 2009). This study was done, however, using the multi-fluid model for neutrals.

The components of the magnetic field are shown in (Figures 8a-b) along V1 (a) and V2 (b). These profiles are expressed in the *RTN* spacecraft frame as in (Figures 7a-b). The *R* axis is defined by the respective (V1) and (V2) heliocentric radius vectors. Because of the nature of the Parker field, the interplanetary field is mainly azimuthal upstream of the TS, i.e. along the *T* direction or $e_\phi$ in spherical coordinates. Hence the overlap between the total field (black curve) and the tangential field (red curve). Similar behavior is seen for the solar wind flow (Figures 7a-b), where upstream of the TS the solar wind velocity is mainly radial. The deviation of the magnetic field from the Parker field takes place at the heliosheath, where the effect of the flow entrainment with the frozen-in field becomes important. At which point, the remaining components of the field, radial ($B_R$) and normal ($B_N$) to the spacecraft, mimic the behavior of the plasma flow ($V_T$ and $V_N$). It is important to mention that the cancelation of the total magnetic field seen in (Figure 8a) along V1 and the absence of such a feature in (Figure 8b) along V2, is not a reconnection feature. This is simply due to the crossing of the current sheet, directed northward in our model, by the extrapolated V1 trajectory. Magnetic reconnection, on the other hand, between the interstellar magnetic field and the interplanetary field can be expected locally at the HP (Swisdak et al. 2009), and can be a good candidate in explaining the 2-3 kHz radio signal near the HP (Gurnett et al. 2003).

The effect of the interplanetary field ($B_{IP}$) on the plasma properties and the distribution of neutrals through the heliosphere is shown in (Figure 9). Properties of ions and neutrals are compared between models, including the results from the (A-I) model (Izmodenov et al. 2005), which assumed ($B_{IP} = 0$). The profiles shown in (Figure 9) are taken along the X axis upwind the *LISM* flow, which passes close the HP stagnation point. The neutrals density, temperature and flow speed profiles ($n_H$, $T_H$, $U_H$) for the (MF-MHD) model are summed over all 4 neutral populations. This is done to facilitate the comparisons with the other two kinetic models. The main feature seen in (Figures 9) is that for neutrals as well as for ions, the changes induced by the differences in models for neutrals, kinetic or multi-fluid, seem to dominate over any interplanetary field effect. This can be seen from the relative agreement between the (K-MHD, $B_{IP} \neq 0$) and the (A-I, $B_{IP} = 0$) models. One needs to keep in mind that the steady state solution of the highly supersonic solar wind interaction with the *LISM* depends mostly on the equilibrium between the solar wind ram pressure and the total *LISM* pressure. The interplanetary field pressure has a lesser role in the established equilibrium between these two winds. The presence of such field, however, is crucial for the pick-up process of newly ionized neutrals inside of the HP and the TS.

The observed decrease in the protons number density, for the K-MHD model, near the HP (Figure 9d) and the subsequent increase outside, is similar to what is seen in (Figure 5). The reduction of solar wind ram pressure due to the increased filtration ratio leads to an unequal shift of both the TS and HP boundaries (Table 1). The non-uniformity in this shift is due to the relative differences in the charge-exchange rates within the heliosheath and inside the TS region. Thus leading to a compression-rarefaction wave structure in the density profile. The reverse tendency is seen in the velocity profile (Figure 9f).

## 5. Conclusion

The kinetic model (K-MHD) of local interstellar hydrogen, more adequate in describing the transport of neutrals through the heliospheres, showed key differences in the heliospheric boundaries and the heliosheath flows with respect to the multi-fluid model (MF-MHD). Most important of these are the decrease of both the heliocentric distance to TS and the heliosheath thickness. The effect of the local interstellar magnetic field was assessed by varying the direction of the field using different combinations of the angles ($\alpha$ = 20°, 30°) and ($\beta$ = 60°, 80°, 90°) with a field intensity of 4.4 µG. The changes in heliospheric features for both kinetic and multi-fluid models revealed greater dependence on the angle ($\alpha$) more than on ($\beta$), making the former easier to constrain from observations.

The analysis of the deflection angle of the heliosheath flow ($\theta$) showed a strong dependence on the neutral model. Lower deflections are achieved when neutrals are kinetic. These changes in ($\theta$) across models, imply that neutrals affect, in non proportional way, the normal ($V_N$) and tangential ($V_T$) velocity component ($\theta = \tan^{-1}(V_N/V_T)$)). This may be an indicator of anisotropy in the kinetic neutrals as they cross the HP. Nevertheless, the asymmetry of the termination shock is found to be the same for both models of neutrals (~6.9-7AU).

The transport of neutrals through the heliosphere and the coupling with the solar wind protons seems to have a greater impact on the global feature of the heliosphere than the interplanetary magnetic field. The absence of the interplanetary field, however, caused a reduction of some of the plasma properties within the TS boundary as shown by the (A-I) model.

**Acknowledgments:**

The authors would like to thank the staff at NASA Ames Research Center for the use of the Columbia supercomputer. This work was supported by the National Science Foundation Career Grant ATM-0747654 and NASA-Voyager Guest Investigator grant NNX07AH20G. D. Alexashov was supported in part by RFBR grants 10-02-93113-НЦНИЛ_a and 10-01-00258-a. V.Izmodenov was supported in part by RFBR grant 10-02-01316-a, Dynastia foundation and the grant of President MD-3890.2009.2.

**Figures caption:**

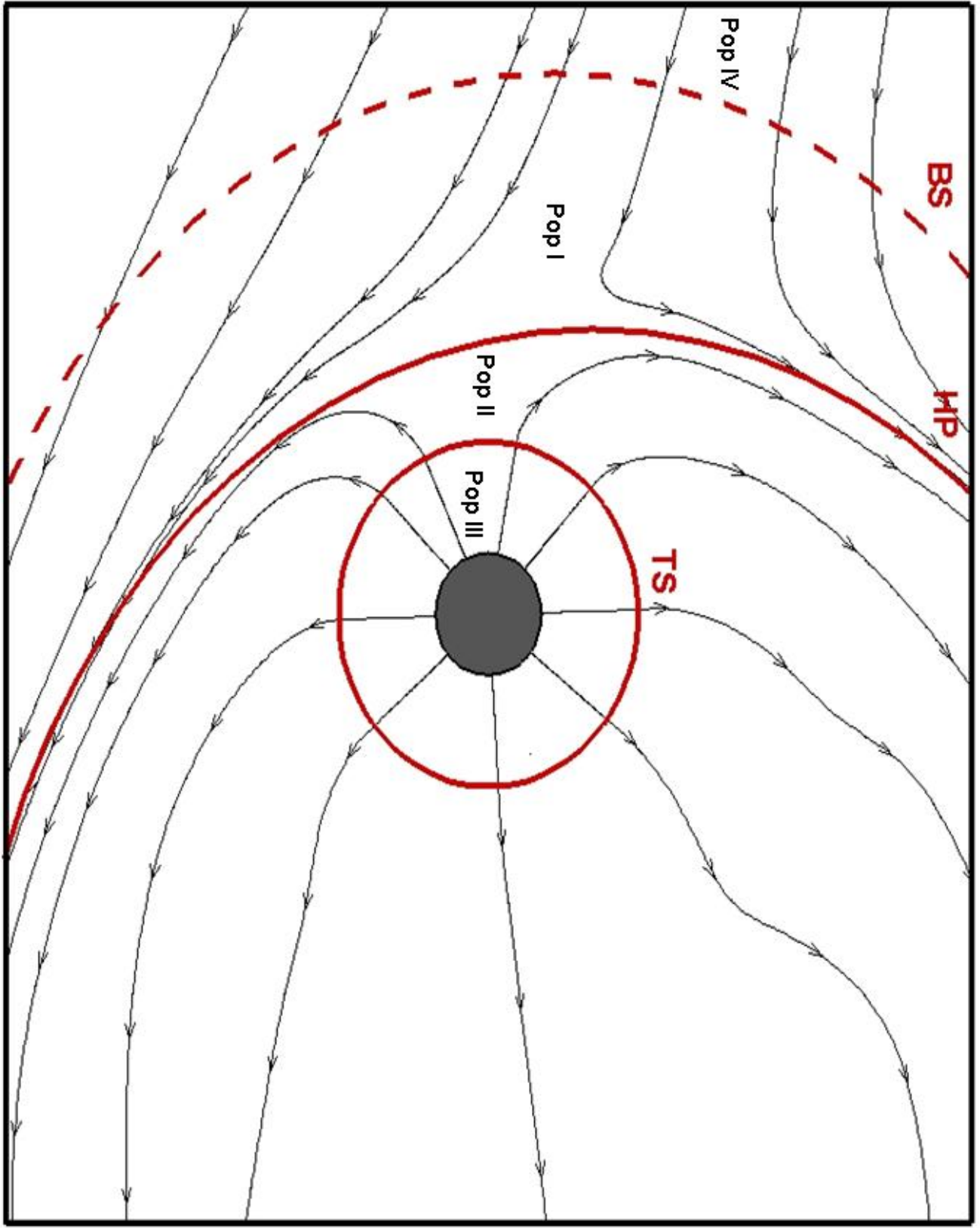

Figure 1: (a) Schematic representation, not to scale, of the heliospheric boundaries and the four neutral populations considered in the multi-fluid model. Numbers (I, II, III, IV) refer to the respective zones of each neutral population. Each neutral population "i" is generated through charge-exchange only in its respective zone "i". When this population charge-exchange in another zone "j", the created neutral population is "j". Here we used the same definitions as used in (Opher *et al*. 2009). (b) The convention chosen for defining the angles (α) and (β). Angle (α) is the angle between $B_{LISM}$ and local interstellar flow velocity $V_{LISM}$; while (β) is the angle between the *BV* plane and the solar equatorial plane respectively.

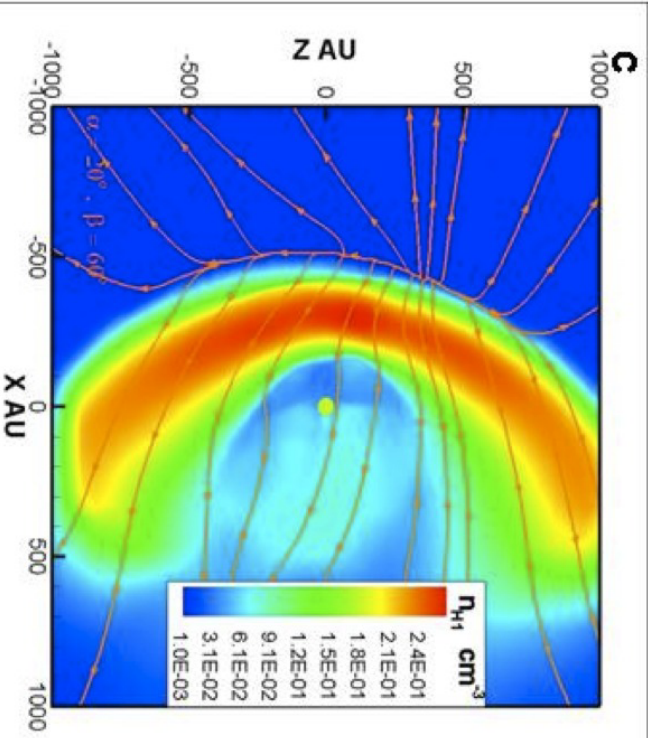
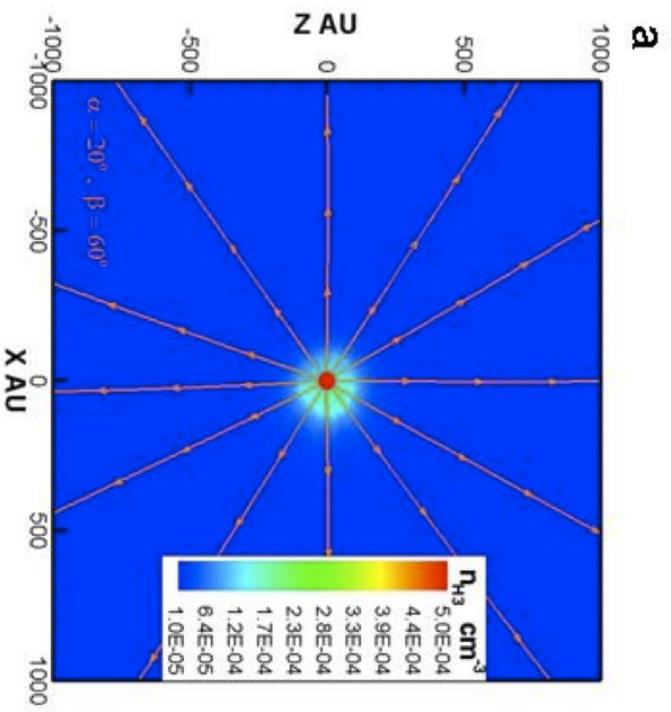
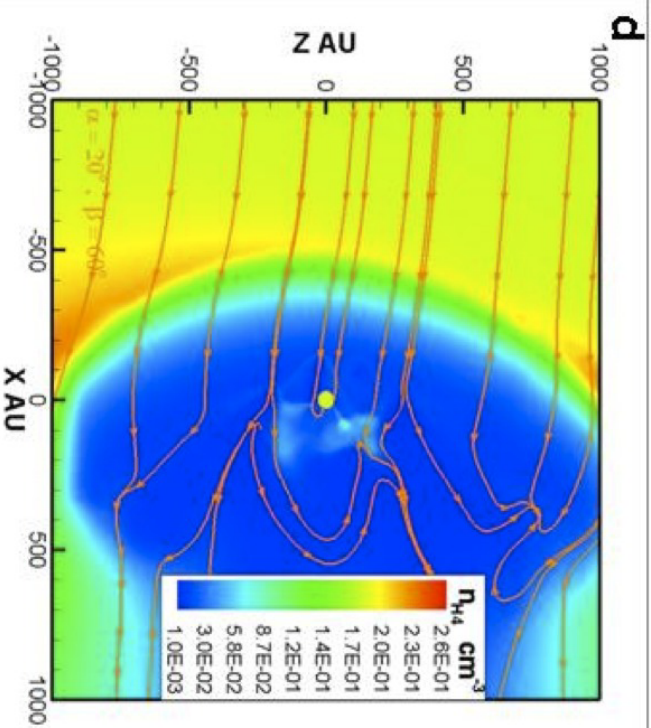
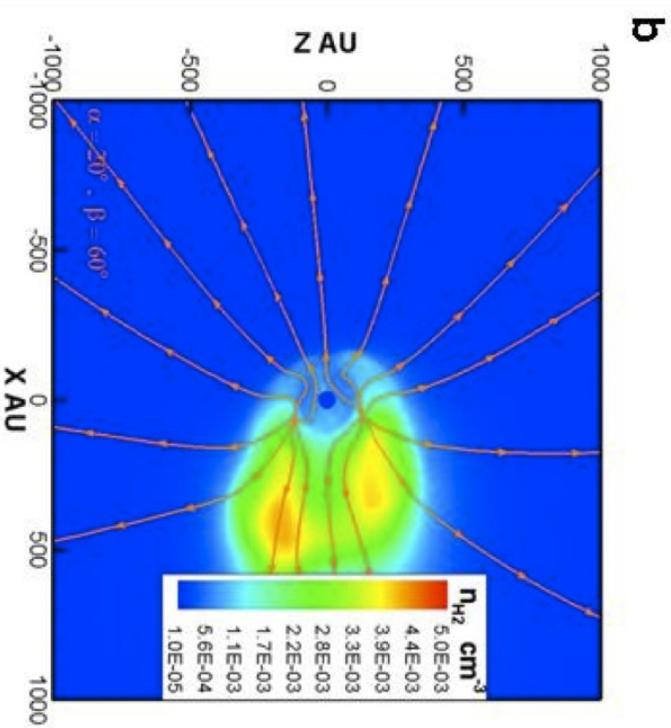

Figure 2a: Density map of neutral hydrogen, multi neutral populations, (MF-MHD) model. The $B_{LISM}$ intensity is 4.4µG and its direction is defined by ($\alpha$ = 20° and $\beta$ = 60°). The stream lines represent the velocity components ($U_x$, $U_z$) of each respective specie.

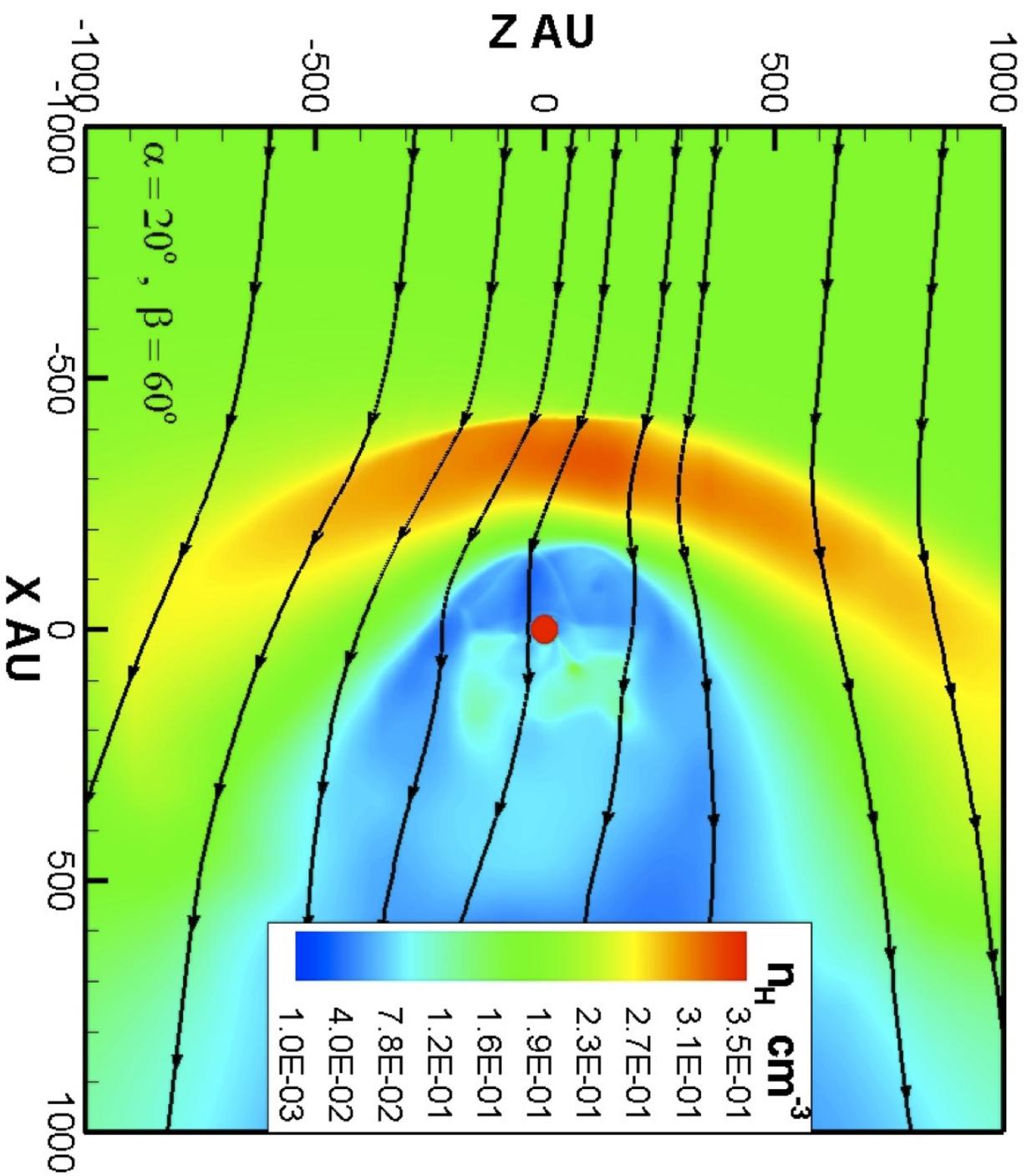

Figure 2b: Density map of neutral hydrogen, the sum of all populations, (MF-MHD) model. The $B_{LISM}$ intensity is 4.4µG and its direction is defined by ($\alpha$ = 20° and $\beta$ = 60°). The stream lines represent the density weighted velocity components ($U_x$, $U_z$) for the sum of all neutral species.

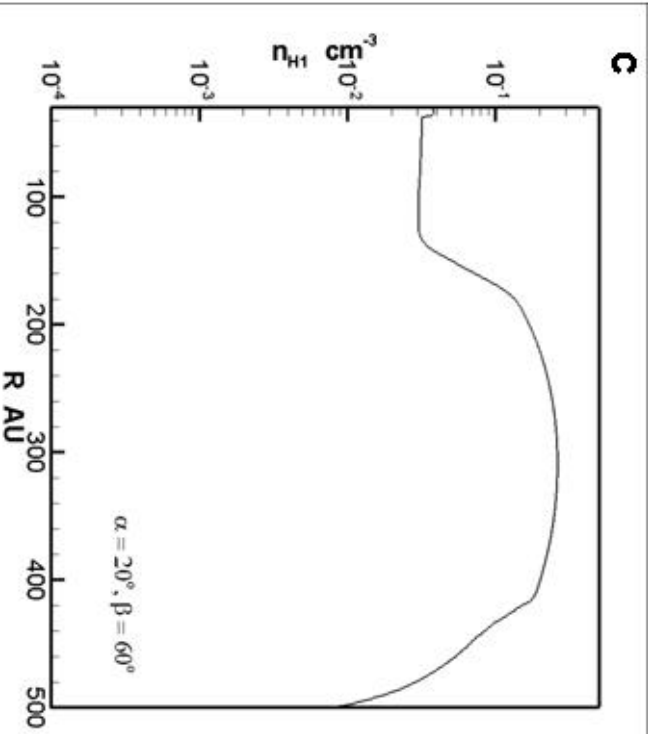
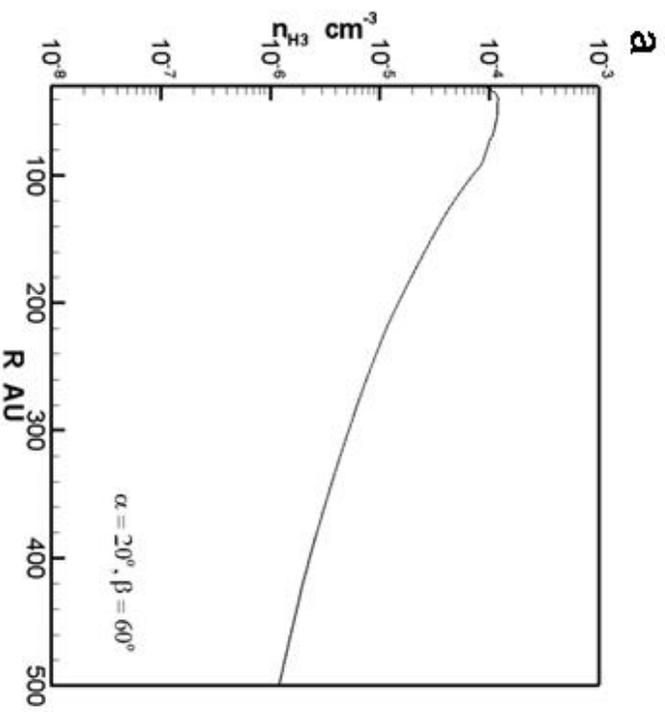
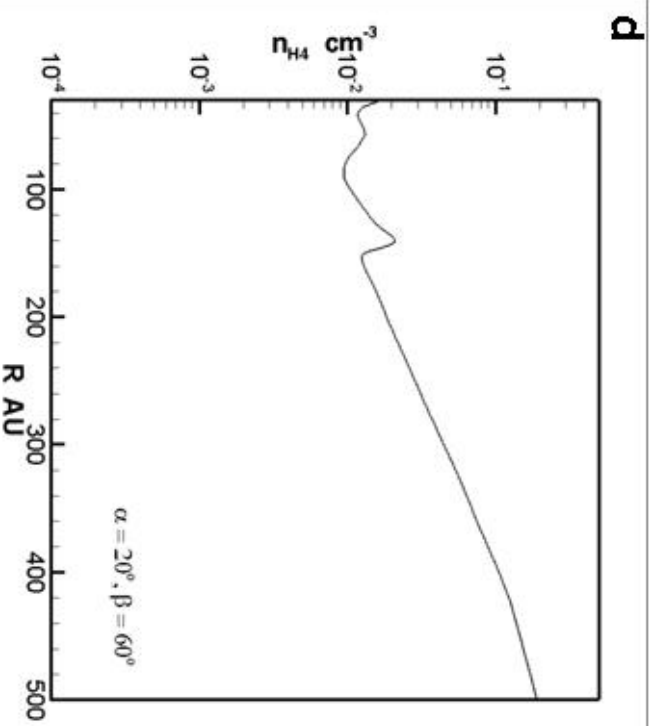
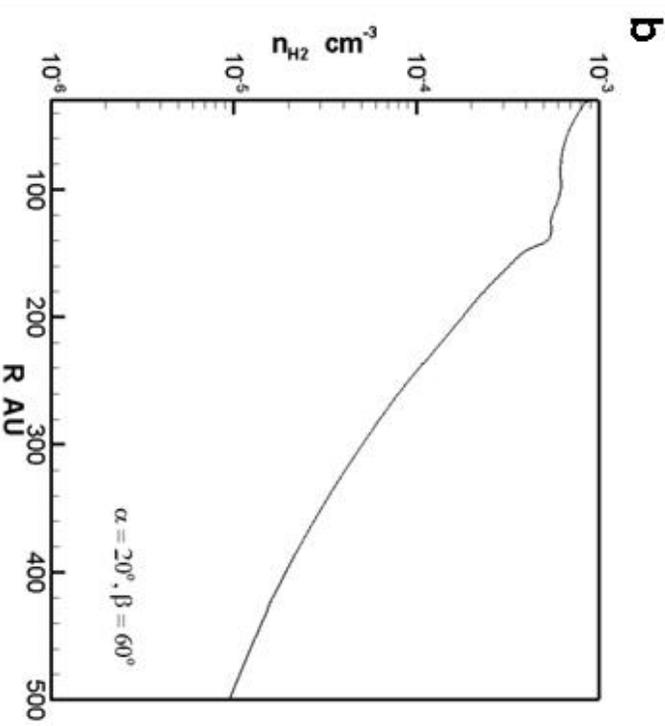

Figure 3a: Radial density profile, for multi neutral populations, along the X axis in the upwind *LISM* flow direction, (MF-MHD) model. The $B_{LISM}$ intensity is 4.4µG and its direction is defined by (α = 20° and β = 60°).

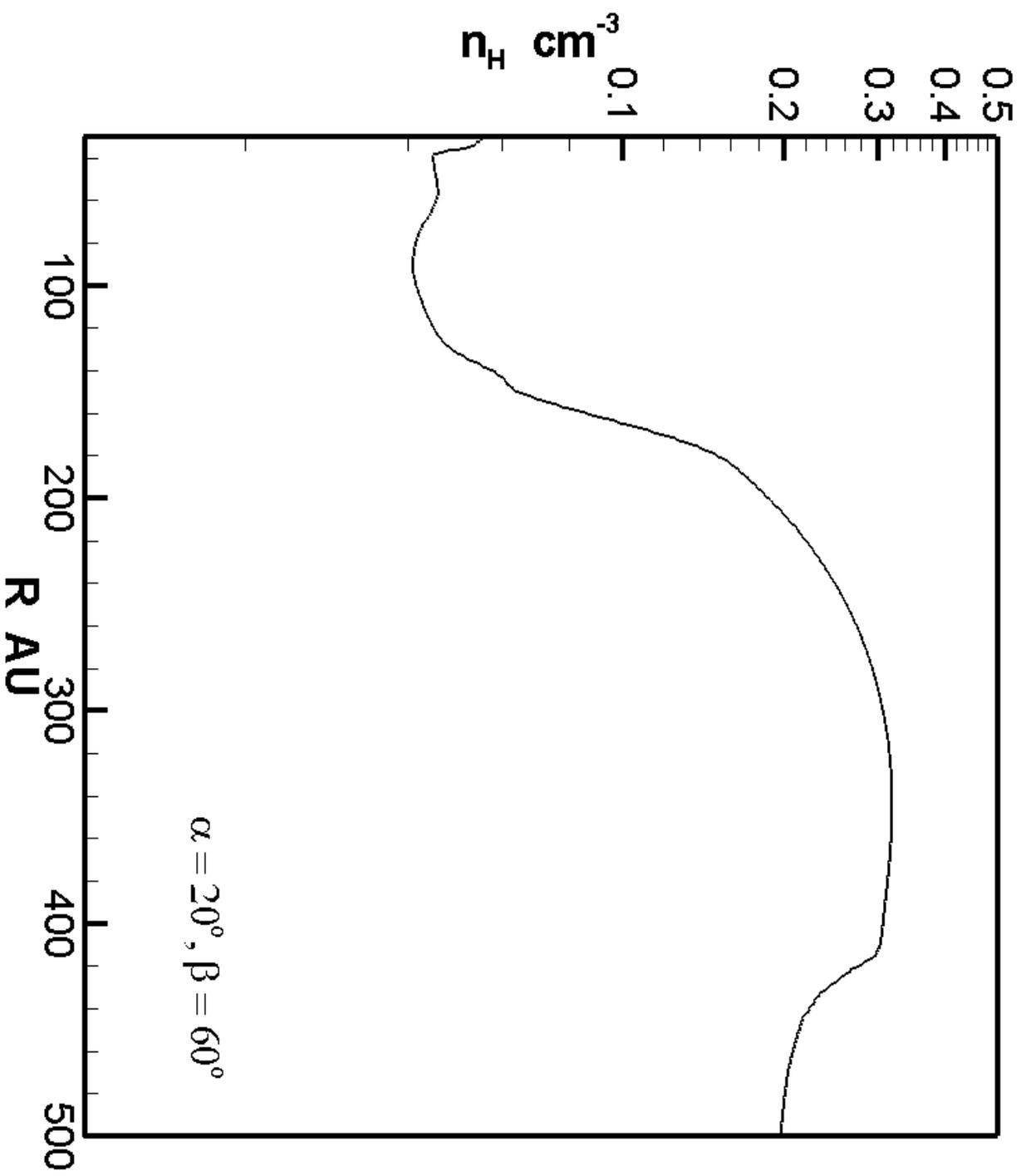

Figure 3b: Radial density profile, for the sum of all populations, along the X axis in the upwind *LISM* flow direction, (MF-MHD) model. The $B_{LISM}$ intensity is 4.4µG and its direction is defined by (α = 20° and β = 60°).

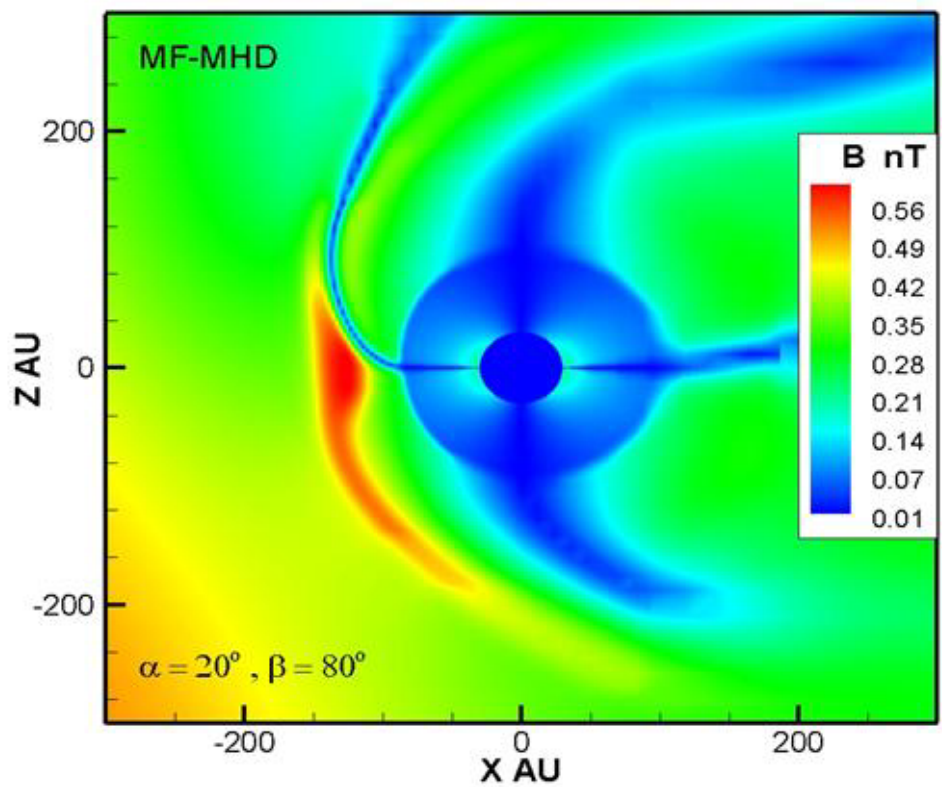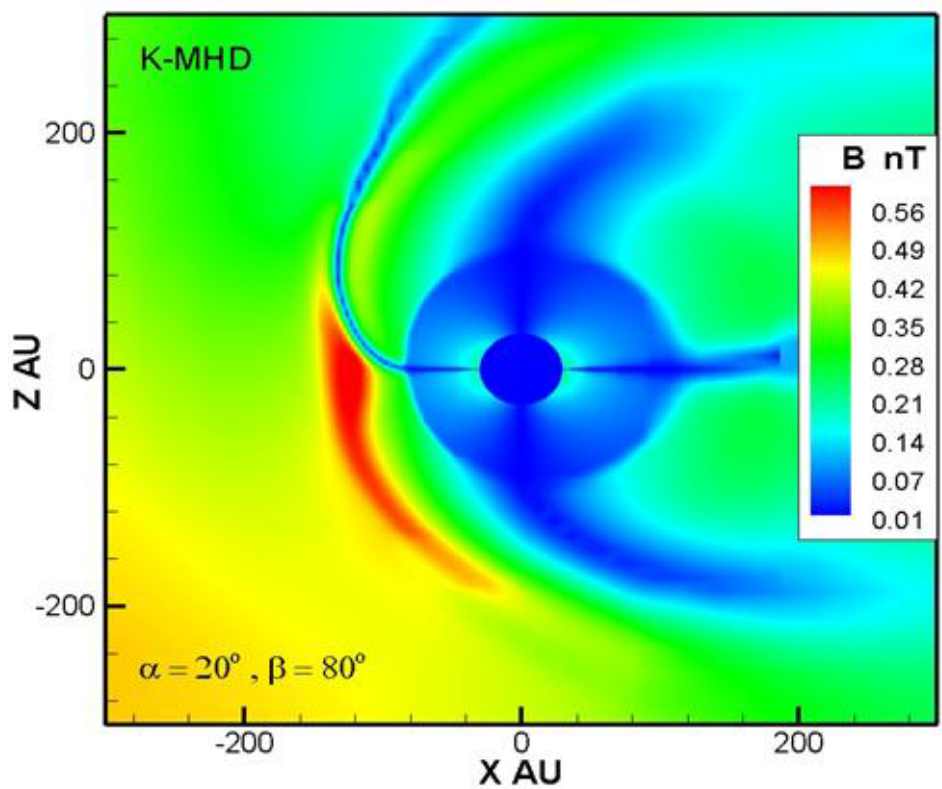

Figure 4: The 2d intensity map of the magnetic field in the XZ plane for the two models (MF-MHD) (a) and (K-MHD) (b). The $B_{LISM}$ intensity is 4.4µG and its direction is defined by ($\alpha = 20°$ and $\beta = 80°$).

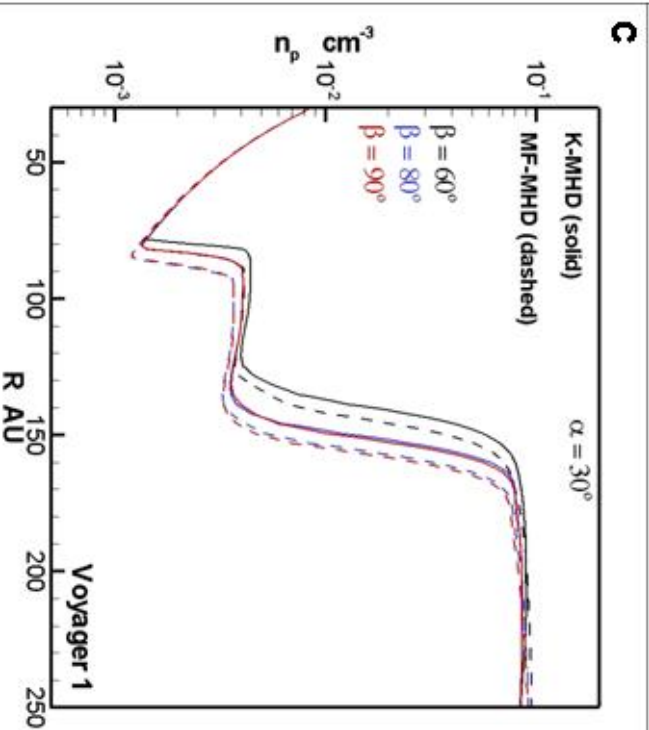
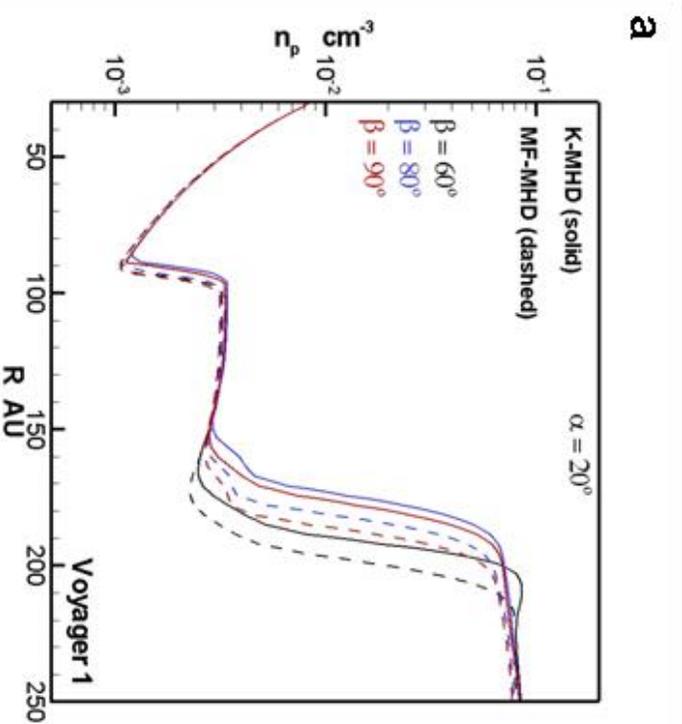
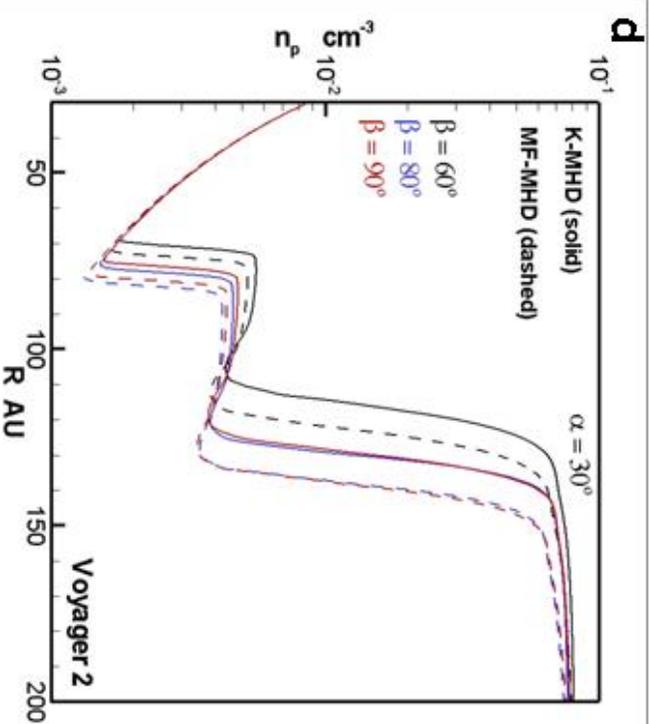
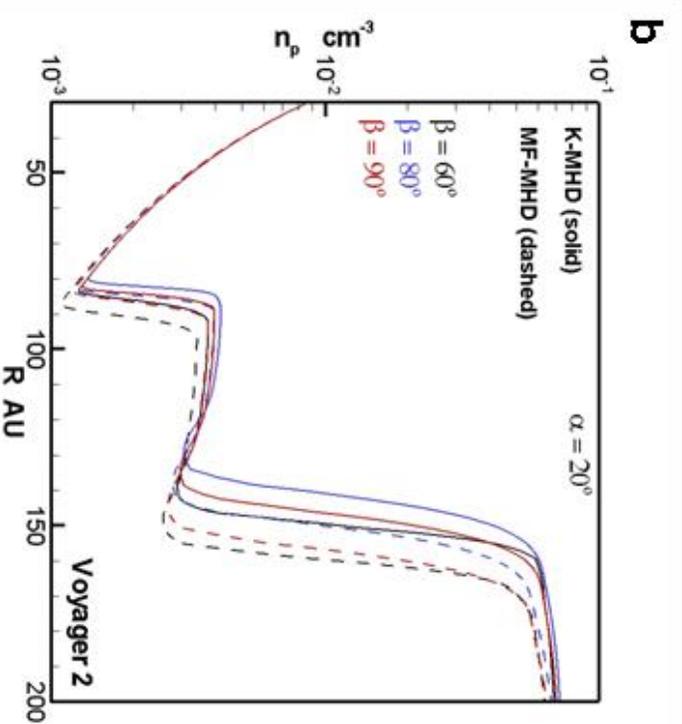

Figure 5: Protons number density profiles along Voyager 1 and Voyager 2 trajectories. Neutral hydrogen is accounted for using two models: multi-fluid (dashed lines) and kinetic (solid lines). The $B_{LISM}$ intensity is 4.4µG. Data along V1 trajectory are shown in the left column, while the right column is for data along V2 trajectory. The upper row is for the cases with ($\alpha = 20°$, $\beta = 60°$, 80°, 90°). The lower row is for ($\alpha = 30°$, $\beta = 60°$, 80°, 90°).

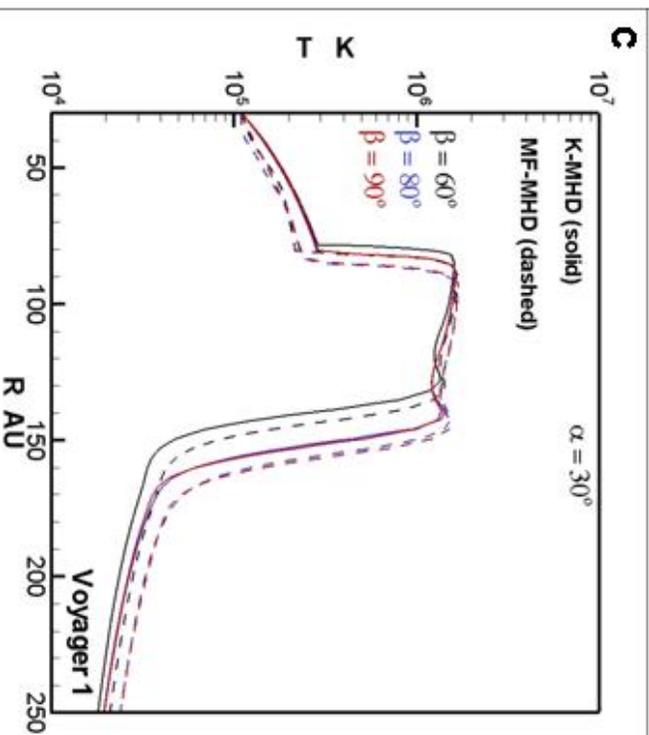
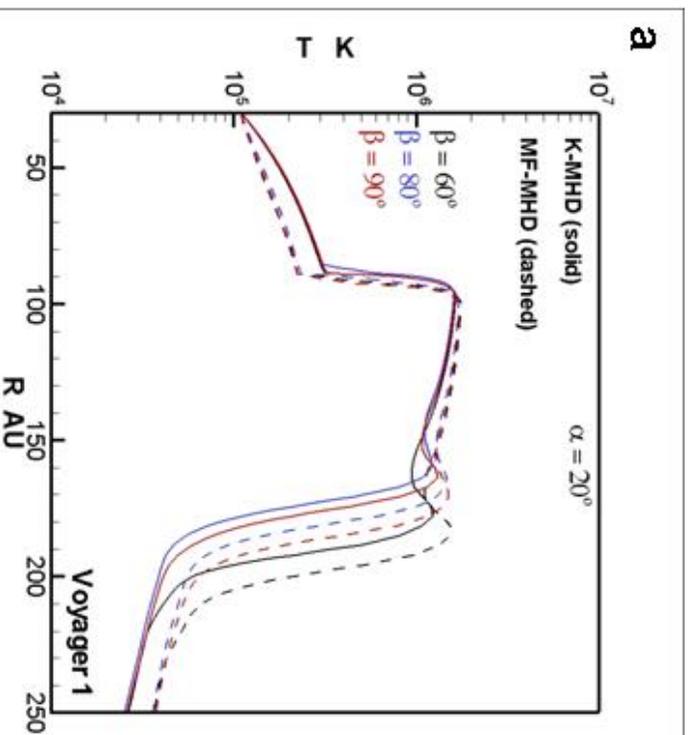
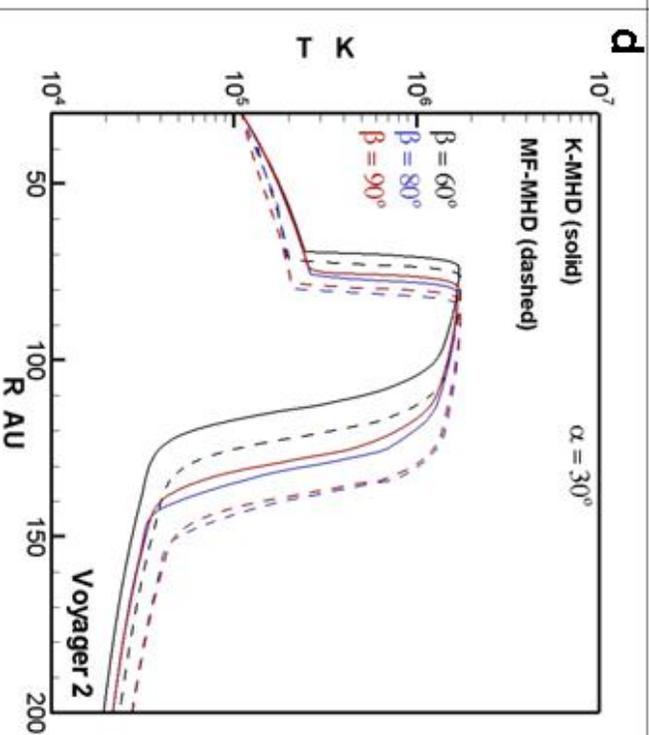
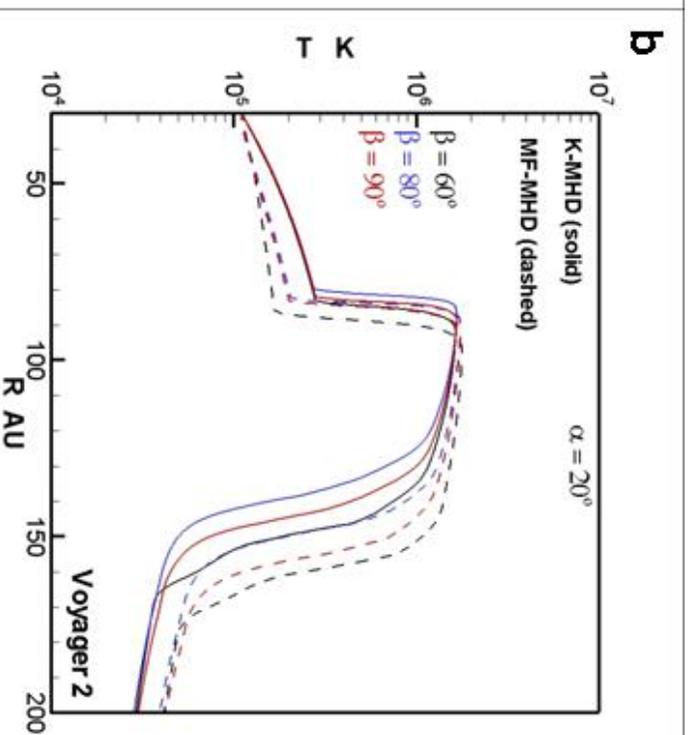

Figure 6: Plasma temperature profiles along Voyager 1 and Voyager 2 trajectories. Neutral hydrogen is accounted for using two models: multi-fluid (dashed lines) and kinetic (solid lines). The $B_{LISM}$ intensity is 4.4µG. The left column is for data along V1, while the right column is for data along V2 trajectory. Data along V1 trajectory are shown in the left column, while the right column is for data along V2 trajectory. The upper row is for the cases with (α = 20°, β = 60°, 80°, 90°). The lower row is for (α = 30°, β = 60°, 80°, 90°).

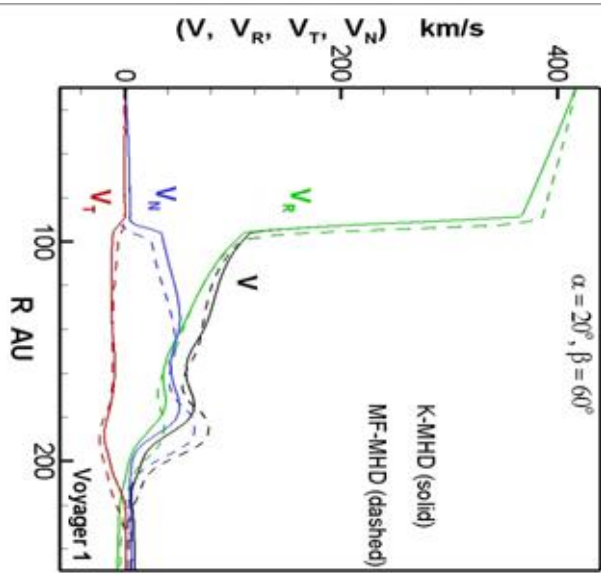
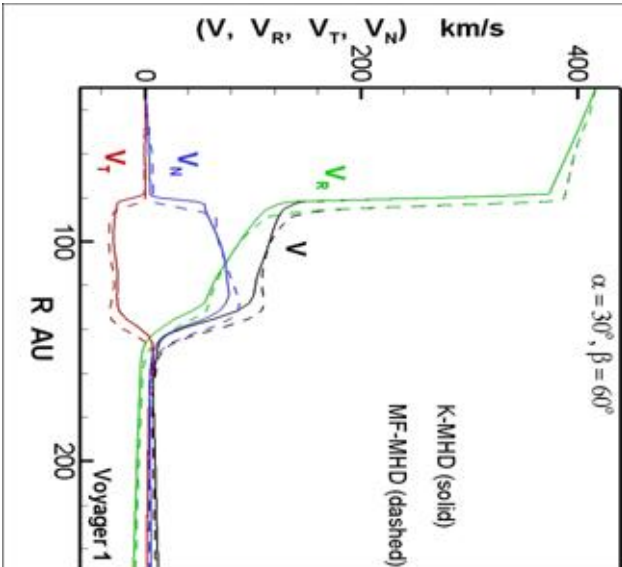
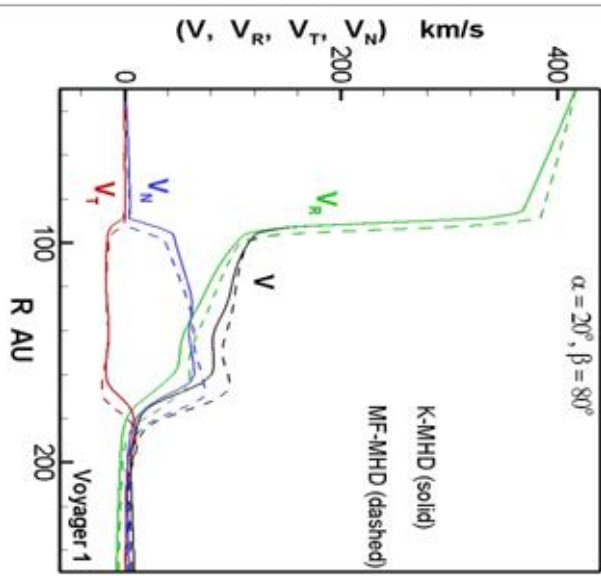
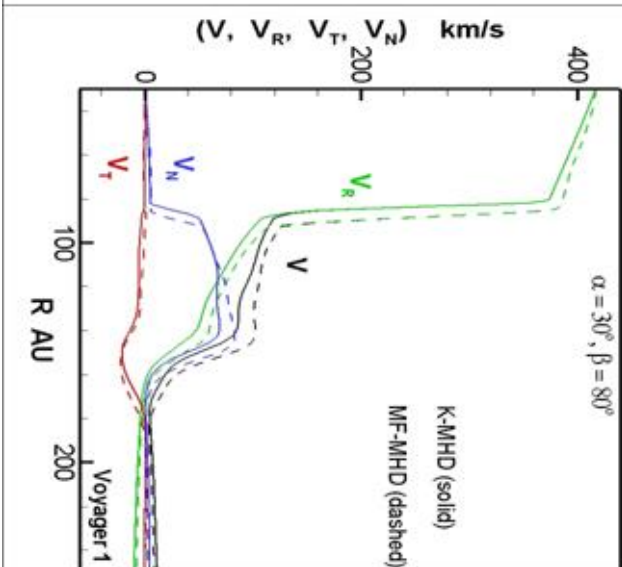
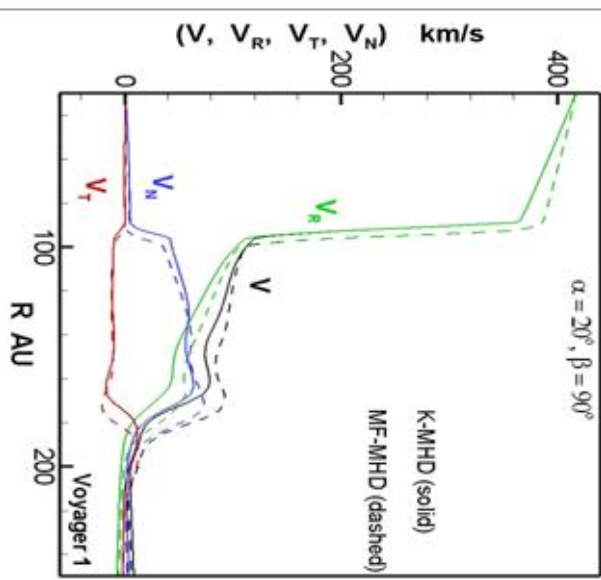
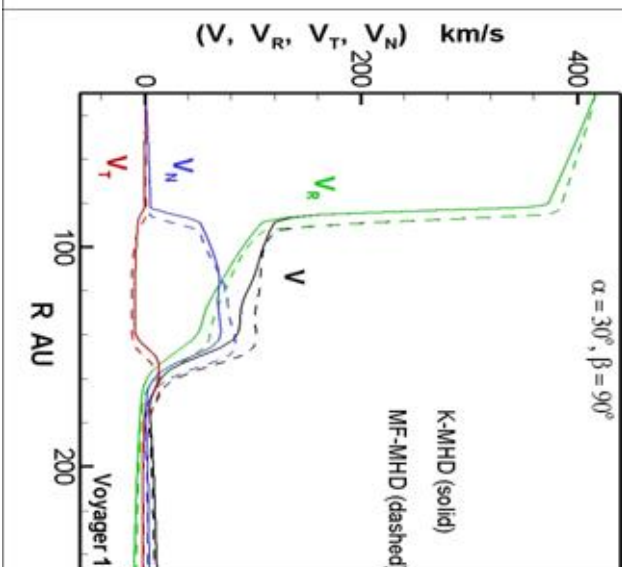

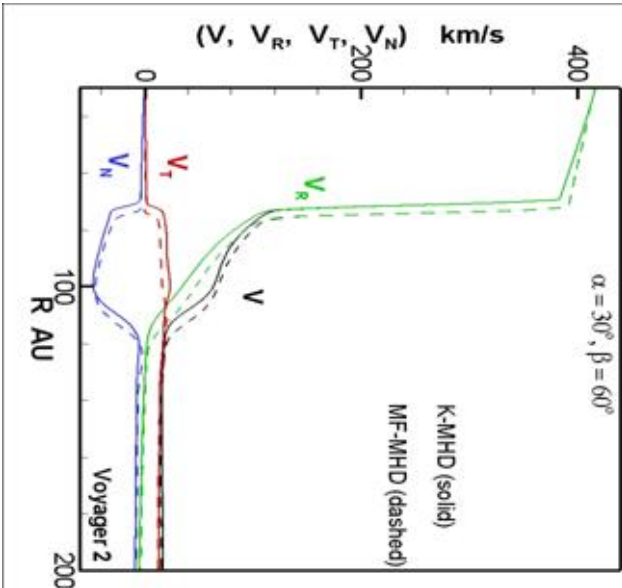
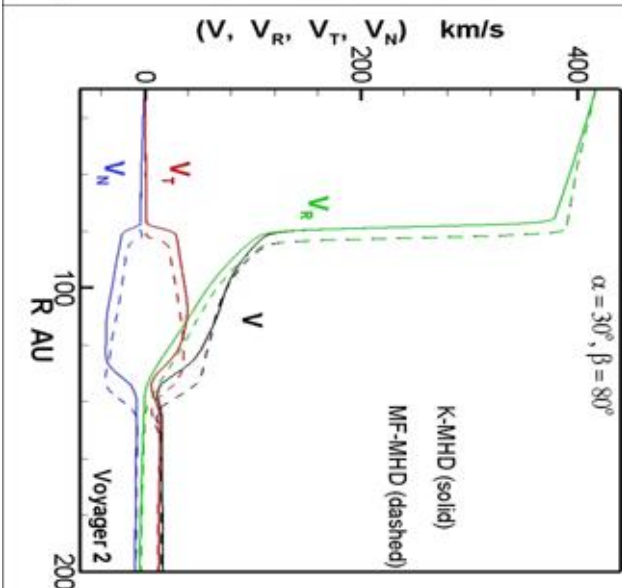
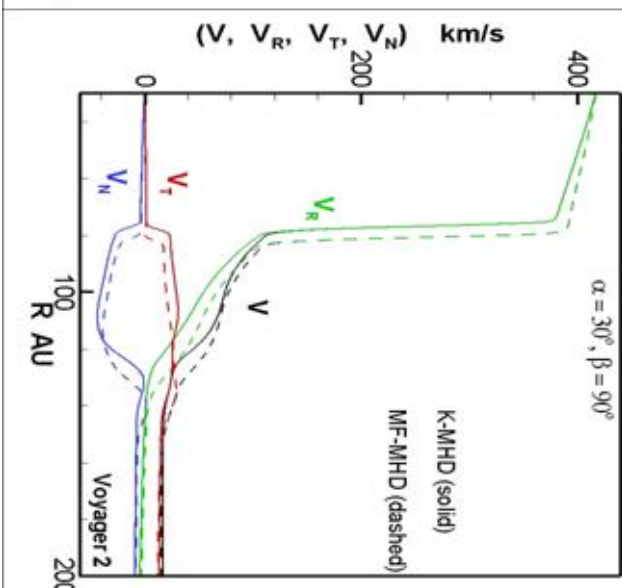
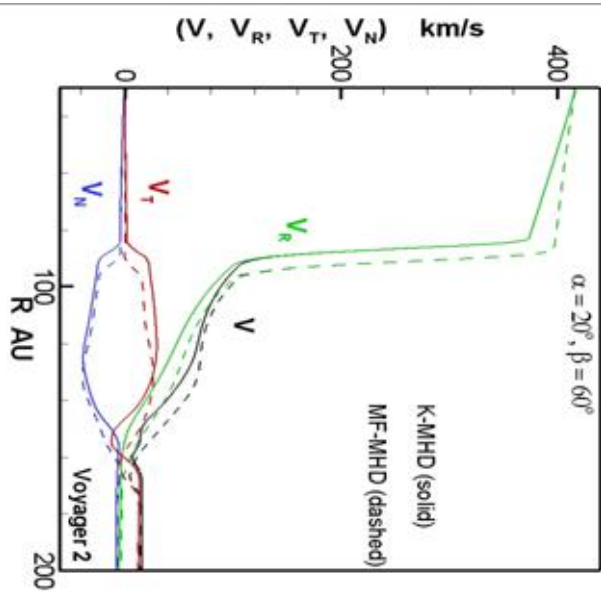
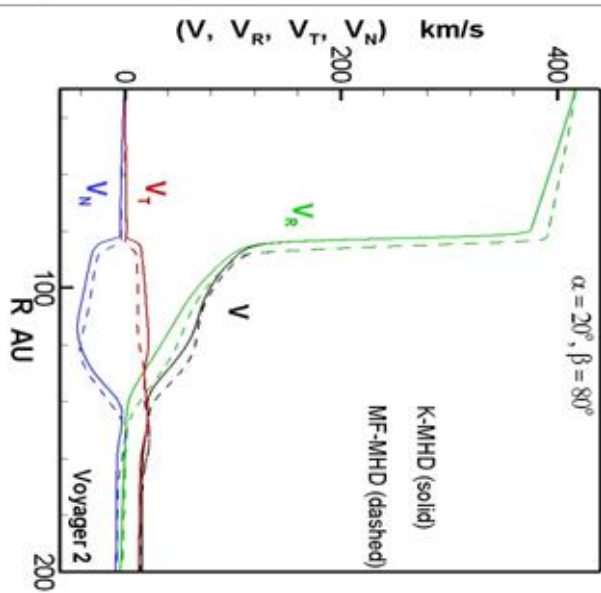
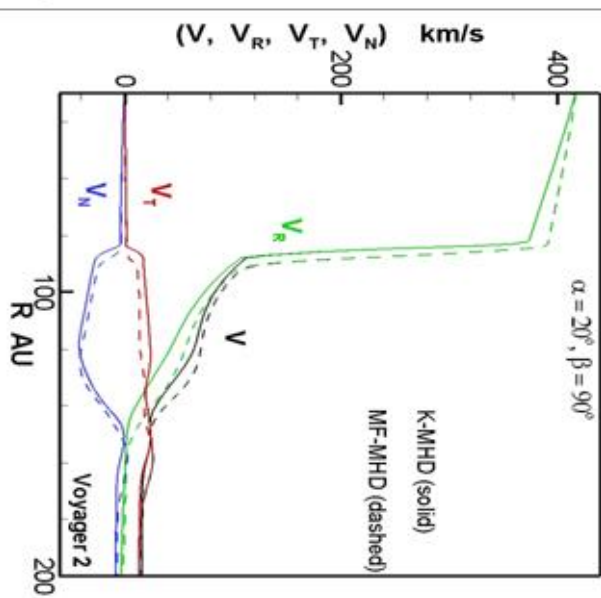

Figure 7: Plasma flow velocities ($V$, $V_R$, $V_T$, $V_N$) along Voyager 1 trajectory; (dashed lines) for the multi-fluid model and (solid lines) for the kinetic model. The $B_{LISM}$ intensity is 4.4µG. The upper row is for the cases with (α=20°), the lower row is for (α=30°). (a) Voyager 1 trajectory. (b) Voyager 2 trajectory.

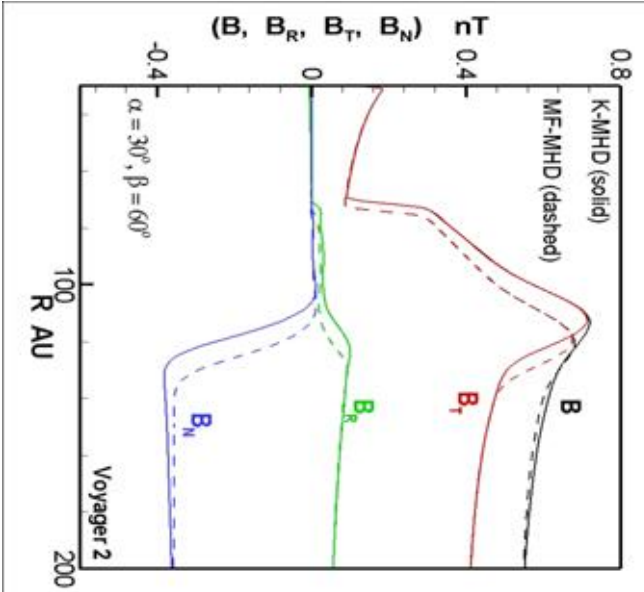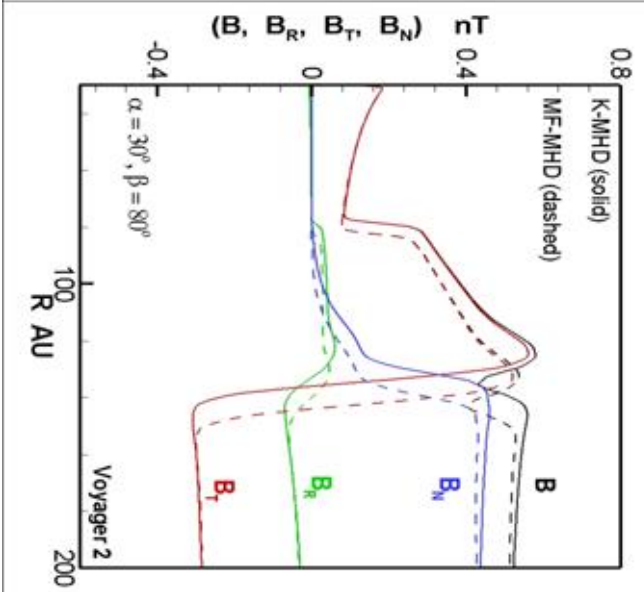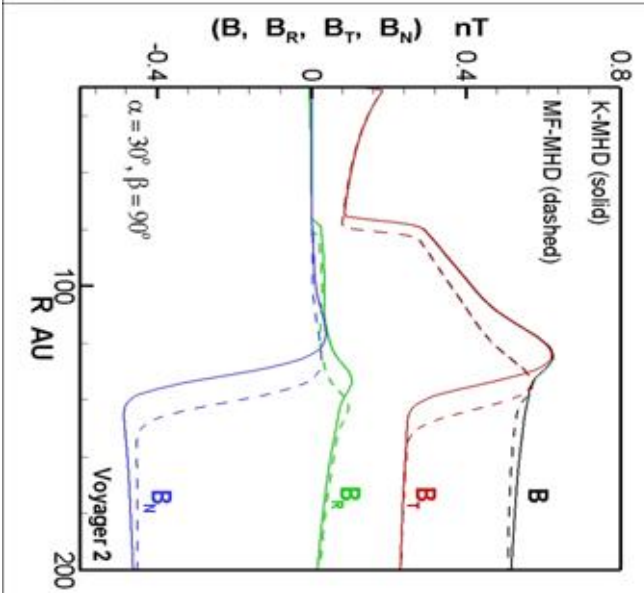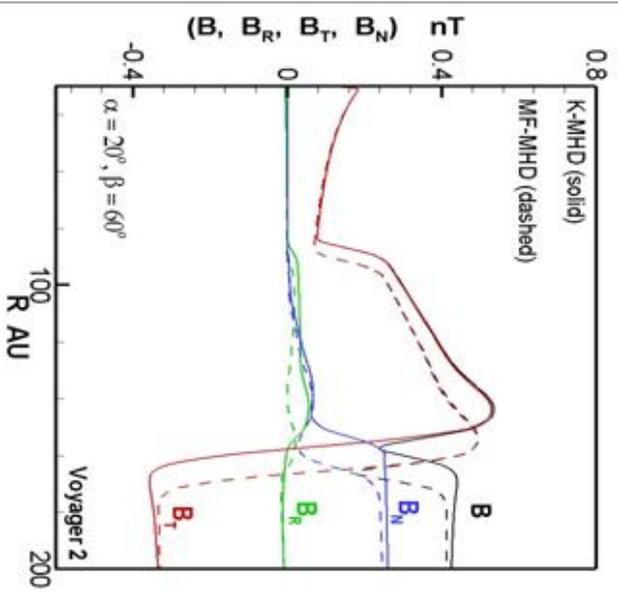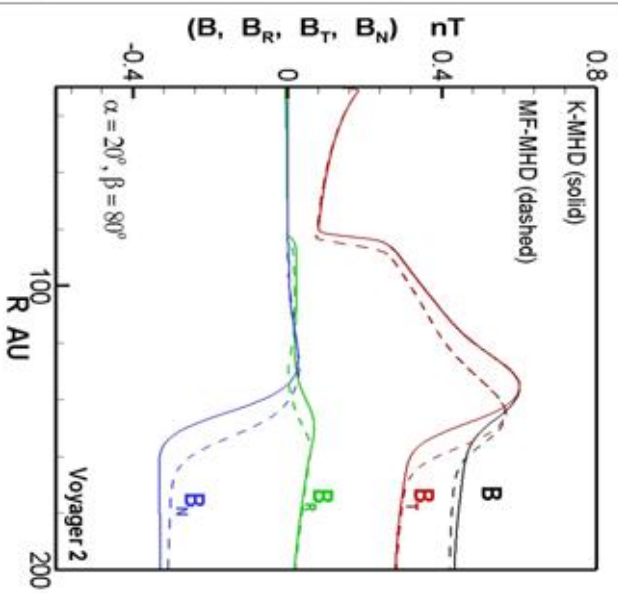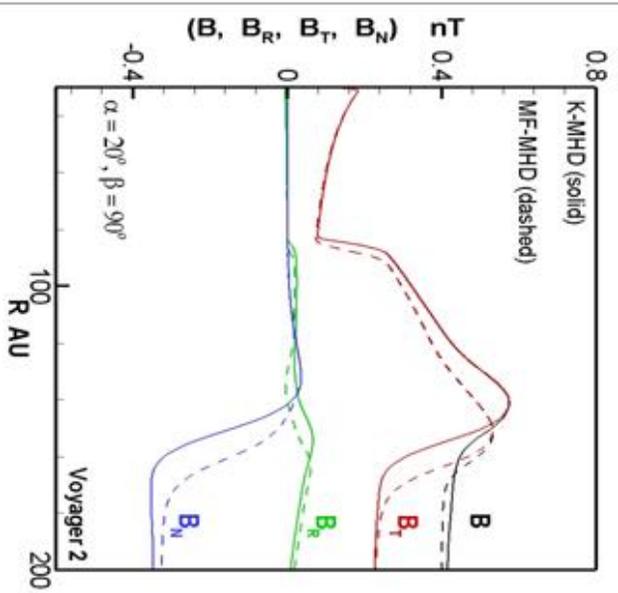

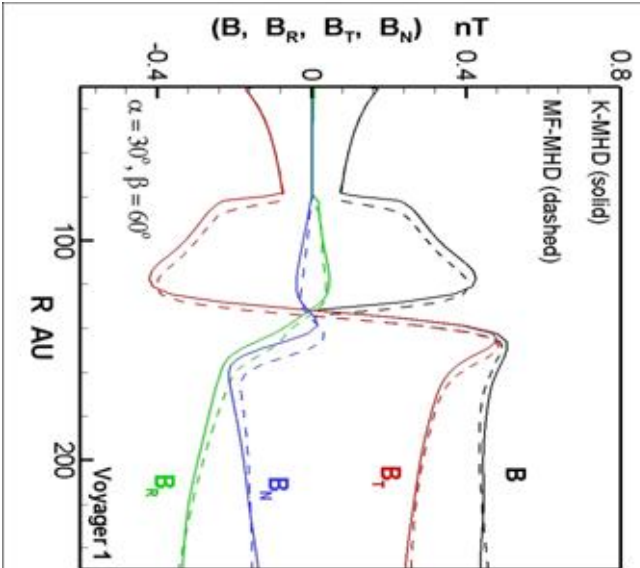
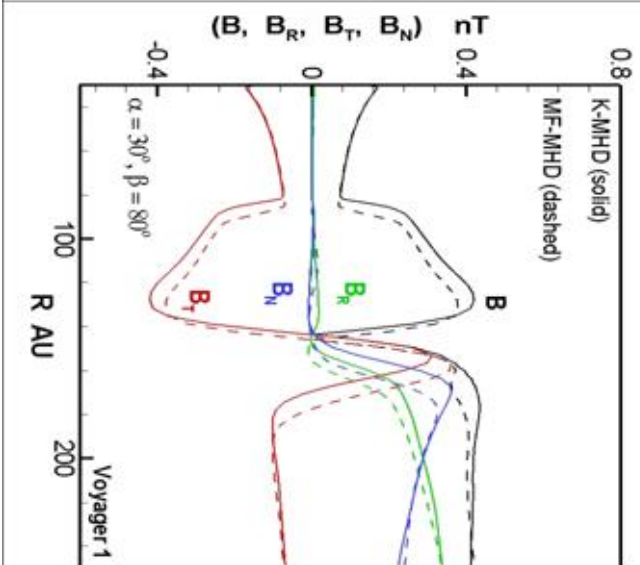
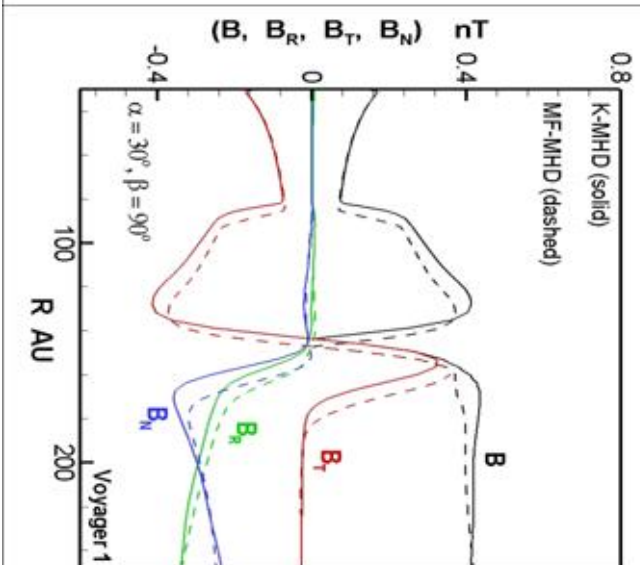
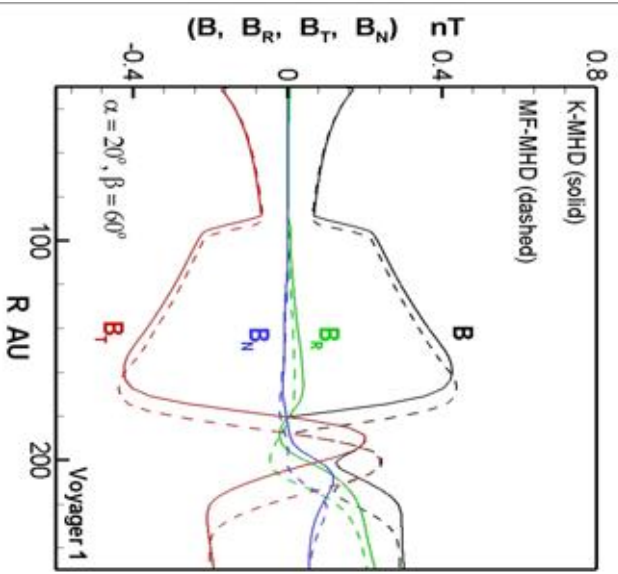
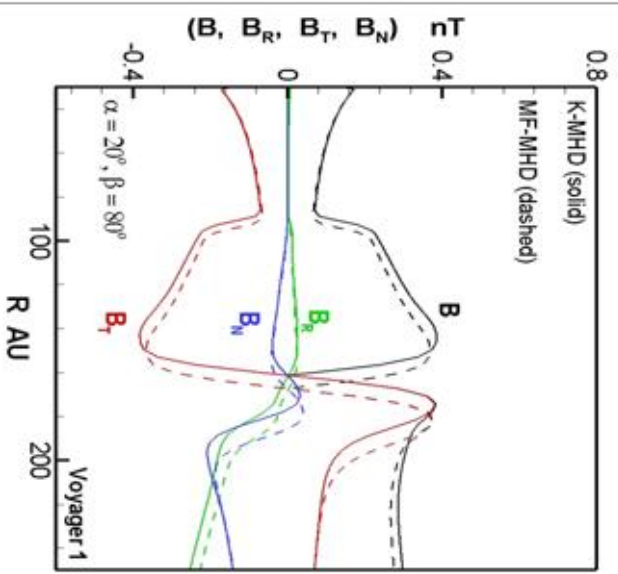
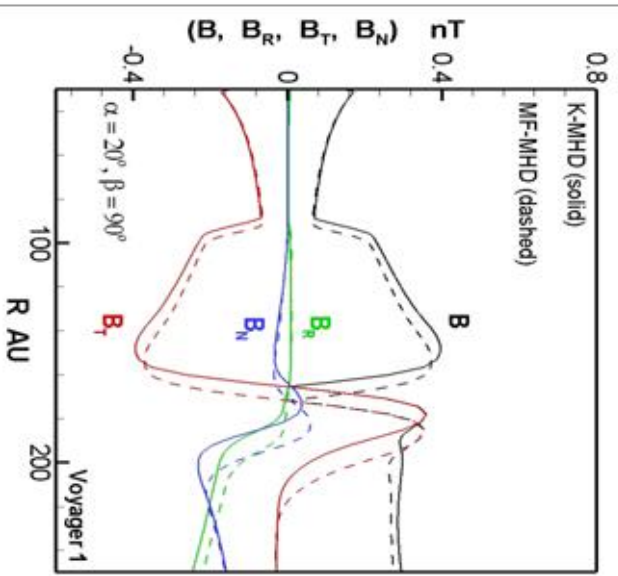

Figure 8: Magnetic field components ($B$, $B_R$, $B_T$, $B_N$) along Voyager 1 trajectory; (dashed lines) for the multi-fluid model and (solid lines) for the kinetic model. The $B_{LISM}$ intensity is 4.4µG. The upper row is for the cases with (α=20°), the lower row is for (α=30°). (a) Voyager 1 trajectory. (b) Voyager 2 trajectory.

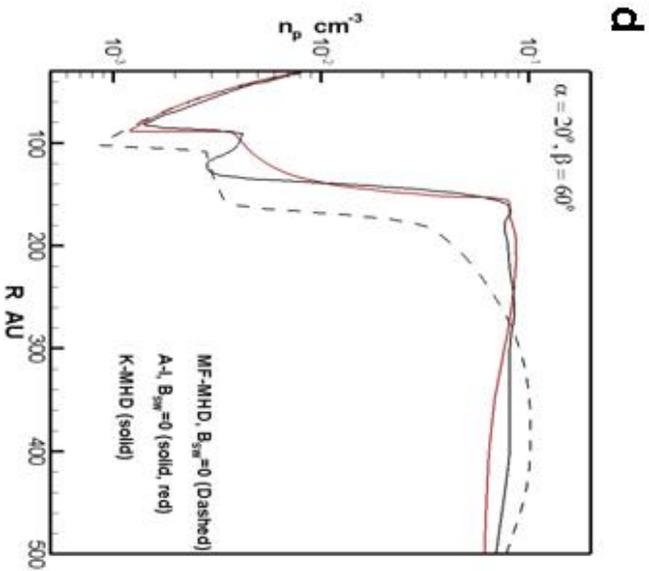
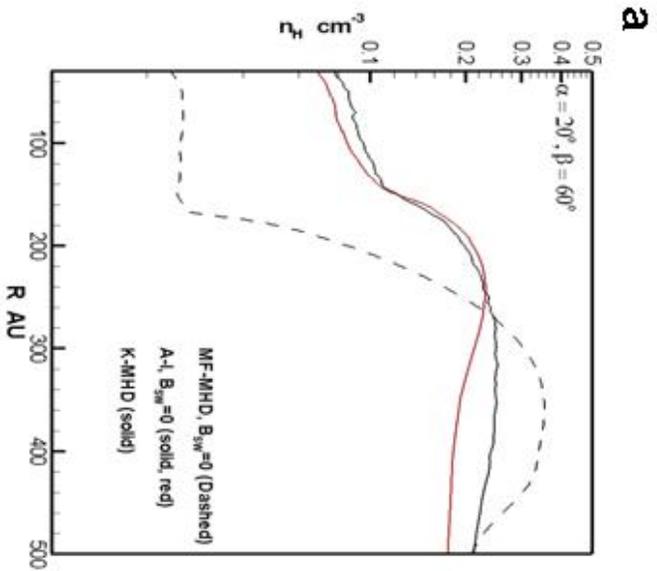
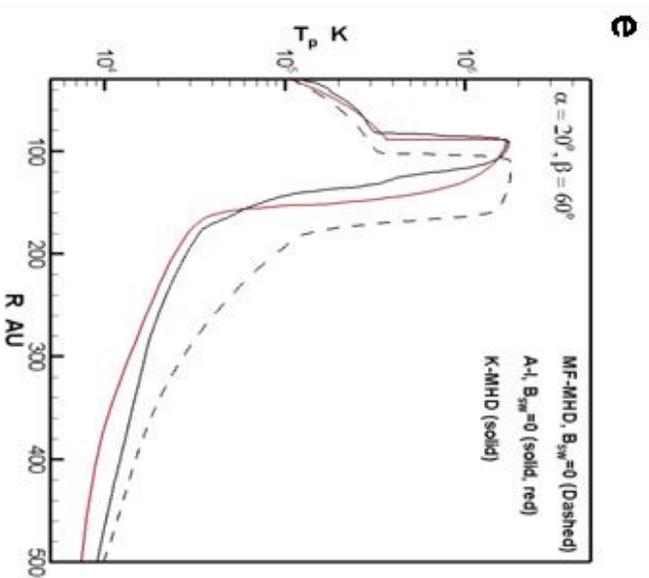
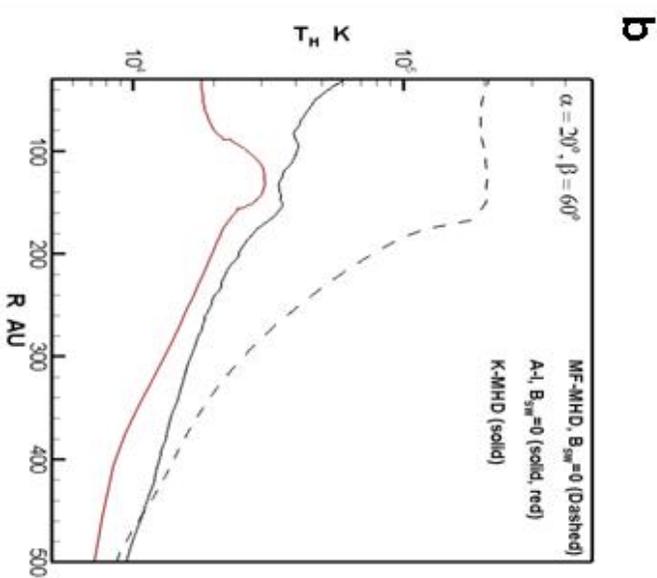
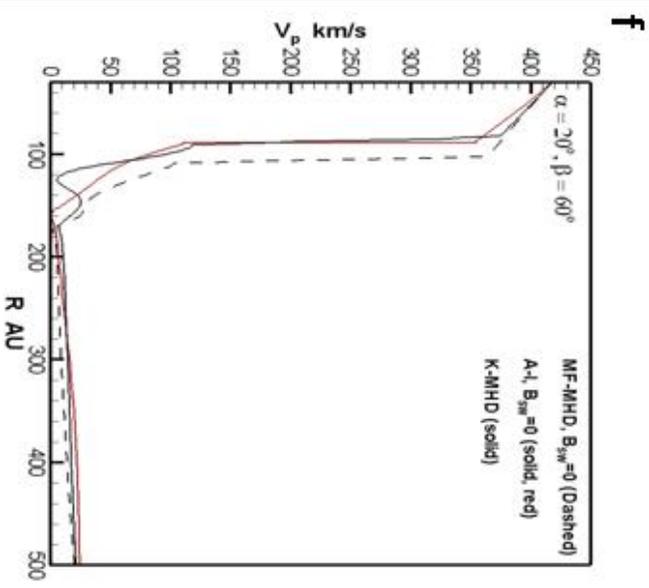
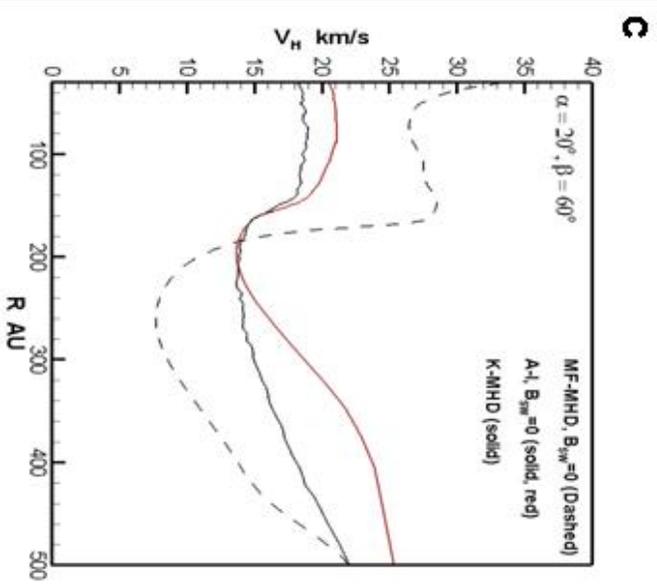

Figure 9: Neutral hydrogen and plasma density (a, d), temperature (b, e) and velocity (c, f) profiles, taken along the X axis in upwind LISM flow direction. Dashed lines represent the MF-MHD model without interplanetary magnetic field. Solid lines (in black) are data from K-MHD model with a Parker solution for the interplanetary magnetic field at the inner-boundary. Solid lines (in red) represent the Moscow University model (A-I) (Izmodenov *et al*. 2005, Alexashov &Izmodenov 2005) without interplanetary magnetic field. The $B_{LISM}$ intensity is 4.4µG, with (α=20° and β=60°).